\documentclass{jfm}
\usepackage{amssymb}
\usepackage{amsmath}
\usepackage{graphicx}
\usepackage{dcolumn}
\usepackage{bm}
\usepackage{epsfig}
\usepackage{float}
\usepackage[T1]{fontenc}
\usepackage{ae,aecompl}
\usepackage{epstopdf, epsfig}
\usepackage{color}
\usepackage{appendix}
\usepackage[export]{adjustbox}
\usepackage{graphicx}
\usepackage{epstopdf, epsfig}
\usepackage[section]{placeins}
\usepackage[latin1]{inputenc}
\usepackage{subfiles}
\usepackage{natbib}
\usepackage{hyperref}
\usepackage{booktabs}
\usepackage{makecell}
\usepackage{subcaption}

\hypersetup{
    colorlinks = true,
    urlcolor   = red,
    linkcolor = blue,
    citecolor = blue,
    filecolor = black,
}
\allowdisplaybreaks[4]
\shorttitle{Supersonic flow kinetics}
\shortauthor{Gan, Zhuang, Yang, Xu, Zhang, Chen, Song, Wu}
\affiliation{\aff{1}Hebei Key Laboratory of Trans-Media Aerial Underwater Vehicle, North
China Institute of Aerospace Engineering, Langfang 065000, China
\aff{2}School of Energy and Safety Engineering, Tianjin Chengjian University, Tianjin 300384,
China
\aff{3}National Key Laboratory of Computational Physics, Institute of Applied Physics and
Computational Mathematics, P. O. Box 8009-26, Beijing 100088, China
\aff{4}National Key Laboratory of Shock Wave and Detonation Physics, Mianyang 621999, China
\aff{5}HEDPS, Center for Applied Physics and Technology, and College of
Engineering, Peking University, Beijing 100871, China
\aff{6}State Key Laboratory of Explosion Science and Technology, Beijing
Institute of Technology, Beijing 100081, China
\aff{7} School of Physical Science and Technology, Guangxi University,
Nanning 530004, China
\aff{8} School of Aeronautics, Shan Dong Jiaotong University, Jinan
250357, China
\aff{9}School of Aerospace Engineering, Beijing Institute of
Technology, Beijing 100081, China
}

\begin{document}

\title{Supersonic flow kinetics: Mesoscale structures, thermodynamic
nonequilibrium effects and entropy production mechanisms}
\author{Yanbiao Gan\aff{1}, Zhaowen Zhuang\aff{1}, Bin Yang\aff{2}, Aiguo Xu\aff{3,4,5,6}, Dejia
Zhang\aff{7}\corresp{\email{zdj202479@gxu.edu.cn}}, Feng Chen\aff{8}%
\corresp{\email{chenfeng-hk@sdjtu.edu.cn}}, Jiahui Song%
\aff{9}, \and Yanhong Wu\aff{1}}
\maketitle

\begin{abstract}
Supersonic flow is a typical nonlinear, nonequilibrium, multiscale, and
complex phenomenon. Compared to scenarios described by Navier-Stokes (NS)
theory, mesoscale behavior in supersonic flow exhibits greater discreteness
and a higher degree of nonequilibrium. Meanwhile, the entropy production
mechanism, which involves compression efficiency and the ability to perform
external work, is a significant focus in supersonic flow fields. As the
degree of discrete/nonequilibrium effects increases, the study of entropy production based on
NS theory may need reevaluation and appropriate revision. To address these
issues, this paper develops and applies the discrete Boltzmann \emph{%
modeling and analysis} method/model (DBM), based on kinetic and mean-field theories, to
simulate, and analyze these phenomena. The study provides the following
results: A Burnett-level DBM suitable for supersonic flow is constructed
based on the Shakhov-BGK model. Higher-order analytical expressions for
typical thermodynamic nonequilibrium effects are derived, providing a
constitutive basis for improving traditional macroscopic hydrodynamics
modeling. Criteria for evaluating the validity of DBM are established by
comparing numerical and analytical solutions of nonequilibrium effects. The
multiscale DBM is used to investigate discrete/nonequilibrium
characteristics and
entropy production mechanisms in shock regular reflection. The findings
include: (a) Compared to the NS-level DBM, the Burnett-level DBM offers more
accurate representations of viscous stress and heat flux, i.e., more precise
dissipative mechanisms, ensures non-negativity of entropy production in
accordance with the second law of thermodynamics, and exhibits better
numerical stability.
(b) Near the interfaces of incident and reflected shock waves, strong
nonequilibrium driving forces lead to prominent nonequilibrium effects. By
monitoring the timing and location of peak nonequilibrium quantities, the
evolution characteristics of incident and reflected shock waves can be
accurately and dynamically tracked. (c) In the intermediate state, the bent
reflected shock and incident shock interface are wider and exhibit lower
nonequilibrium intensities compared to their final state. This observation
can be used to track the position and straightening process of the bent
shock. (d) The Mach number enhances various kinds of nonequilibrium
intensities in a power-law manner $D_{m,n} \sim \mathtt{Ma}^\alpha $. The
power exponent $\alpha $ and the order of nonequilibrium effects (kinetic
modes) $m$ follows a logarithmic relation $\alpha \sim \ln (m-m_0)$. This
research provides new perspectives and kinetic insights into supersonic flow
studies.
\end{abstract}

\begin{keywords} supersonic flow,
discrete Boltzmann method, mesoscale structure, thermodynamic nonequilibrium effect, entropy
production mechanism
\end{keywords}

\section{Introduction}

\label{sec1}

Supersonic flow, characterized by nonlinearity, nonequilibrium, and
multiscale effects~\citep{pham1989nonequilibrium}, is widely observed in
nature, scientific experiments, and defense engineering. It is significant
not only in fundamental scientific research but also plays a crucial role in
various key national strategic areas. For instance, in the energy sector,
supersonic flow enhances both combustion efficiency and energy conversion
efficiency~\citep{law2010combustion}. In aerospace engineering, supersonic
technology is crucial for achieving high-speed flight and deep-space
exploration~\citep{wang2023unsteady,wang2024experimental,wang2024unsteady}.
In inertial confinement fusion, isotropic supersonic flow serves as a key
mechanism for achieving high-efficiency fusion reactions~%
\citep{zhou2017rayleigh1,zhou2017rayleigh2}.

Supersonic flow exhibits diverse mesoscale structures~%
\citep{huang2018mesoscience} and kinetic modes that vary spatially and
temporally. The interaction between these structures and modes results in
complex, significant, and persistent hydrodynamic nonequilibrium (HNE) and
thermodynamic nonequilibrium (TNE) effects and behaviors~%
\citep{xu-Book,xu2024advances}. For instance, during re-entry into the upper
atmosphere, a spacecraft transitions through the rarefied free-molecular
flow regime, transitional flow regime, and slip flow regime, ultimately
entering a relatively dense continuous flow regime at very high Mach numbers~%
\citep{li2009gas,li2015rarefied}.
significant cross-scale rarefied gas effects.
The gas properties differ completely across each flow regime, with flow
states exhibiting significant cross-scale rarefied gas effects. The
continuous transition between different flow mechanisms requires the model
to have cross-regime adaptability, both in terms of Knudsen number and the
degree of nonequilibrium. Moreover, due to the interaction between
supersonic flow and aircraft, various mesoscale structures and TNE behaviors
exist around supersonic vehicles, such as bow shocks, shock/boundary layer
interactions, shear layers (instabilities, transitions, impacts,
reattachment), viscous disturbances, turbulent transitions, shock
disturbances, flow separation, radiation, dissociation/ionization, and
surface ablation. In near-space hypersonic vehicles, shock reflection and
wave interference occur in the scramjet engine inlet duct and on both large
components and small local objects of the aircraft, resulting in complex
mesoscale structures such as incident shocks, reflected shocks, detached
shocks, reattached shocks, and separation bubbles~%
\citep{chang2017recent,huang2020recent,sawant2022synchronisation,bao2022study}%
.

These mesoscale structures and behaviors in supersonic flow exhibit both
increased discreteness and a higher degree of nonequilibrium, which notably
influence the system's evolutionary trajectory.
prominent cross-scale rarefied gas effects
This is because: (1) near strong discontinuities like shock waves, local
characteristic scales decrease sharply, amplifying discrete effects and
spatial multiscale phenomena; (2) the highly transient, strongly coupled,
and unsteady nature of supersonic flows manifests that the system's flow
characteristic time is no longer negligible compared to thermodynamic
relaxation time, preventing timely relaxation to thermodynamic equilibrium.
Additionally, these mesoscale structures and behaviors are closely
correlated with entropy production~\citep{zhang2016kinetic}, a critical
parameter for evaluating aerodynamic drag, compression efficiency, and the
ability to perform external work in supersonic flow~\citep{song2023}. As the
levels of nonequilibrium and discreteness increase, traditional entropy
production studies based on Navier-Stokes (NS) theory may require
reassessment and approximate revision. Recent studies~%
\citep{rinderknecht2018kinetic,cai2021hybrid,lianqiang2020research}
highlight that mesoscale kinetic phenomena, marked by significant
nonequilibrium effects and discreteness, have a substantial impact on the
success of ignition in ICF. 
Therefore, developing accurate, reliable, and efficient cross-scale models
to quantitatively and dynamically characterize the evolution of mesoscale
structures, nonequilibrium effects and entropy production mechanisms,
clarify the formation mechanisms and evolution laws of these structures, and
understand their impact on flow, heat transfer behaviors, is crucial for the
cross-scale regulation of complex supersonic flow behaviors.

The primary approaches to studying nonequilibrium effects in supersonic
flows include theoretical analysis, engineering experiments, and numerical
simulations~\citep{ChenJianqiang}. Theoretical analysis involves solving and
analyzing nonlinear equations that describe supersonic flow, such as the
Euler, NS, Burnett, and super-Burnett equations. This approach often
necessitates making numerous unreasonable simplifying assumptions, and most
analytical solutions are confined to low-dimensional cases. Additionally,
the strong nonequilibrium effects in supersonic flow lead to significant
nonlinear coupling between multiple physical fields, further challenging the
validity of these equations. The extreme complexity of flow configurations
at supersonic flows also poses substantial challenges to theoretical
analysis.

Engineering experiments, including flight and ground wind tunnel tests, are
essential for studying nonequilibrium effects in supersonic flow~%
\citep{JIANG20203027,gu2020capabilities}. While flight tests provide
first-hand data and directly validate the feasibility of aircraft designs,
they are both costly and technically challenging. Given the highly complex
supersonic flow conditions in actual flight, ground wind tunnel tests are
often replicate only certain aspects of these conditions. As a result, many
ground tests cannot accurately simulate effects caused by high Reynolds
number inflows, rarefied gas dynamics, and high-enthalpy conditions. This
discrepancy often leads to a mismatch between flow states observed in ground
tests and those in actual flight, significantly reducing the value of
ground-based experiments.

Even with the successful execution of the aforementioned tests and
experiments, significant challenges remain in measuring and analyzing
nonequilibrium effects in supersonic flows. \emph{(1) Extremes of
spatiotemporal scales and limitations of measurement equipment}:
Nonequilibrium effects typically manifest over very short timescales and
within extremely small spatial regions. For example, rapid changes in
temperature, density, and pressure across shock waves can occur within
nanoseconds and be confined to micrometer-sized areas. These extreme
temporal and spatial scales impose stringent requirements on the response
time and resolution of experimental equipment. Traditional measurement
instruments, such as pressure probes and thermocouples, often fail to
accurately capture these changes within such short timescales and small
regions, leading to imprecise or erroneous measurements. Optical measurement
techniques~\citep{danehy2015molecular}, while offering higher spatiotemporal
resolution, often rely on the transparency and uniformity of the flow
medium. In the complex environment of supersonic flows, measurement accuracy
may be constrained. Furthermore, non-contact measurement methods, such as
laser interferometry and particle image velocimetry, still encounter
difficulties in capturing rapidly changing small-scale nonequilibrium
phenomena due to high signal noise and complex data processing. \emph{(2)
Complexity of data interpretation and signal processing}: Even if
experimental data are obtained, interpreting them remains a significant
challenge. Nonequilibrium effects are often accompanied by strong signal
fluctuations and complex background noise, making it difficult for
traditional data processing methods to extract meaningful information.
Furthermore, since nonequilibrium phenomena involve multiscale and
multiphysics coupling, data interpretation typically requires the
integration of various theoretical models and numerical simulations, further
increasing the complexity of data analysis. \emph{Currently, the effective
observation and quantitative analysis of these small-scale structures and
fast modes remain key challenges in experimental fluid mechanics.}
Consequently, the complexity of TNE effects and entropy production
mechanisms within such flows remains poorly understood to date.

Compared to engineering tests, numerical simulations offer advantages such
as lower cost, high repeatability, greater efficiency, reduced risk,
adjustable parameters, comprehensive flow field information, and ease of
analysis, making them essential tools for studying nonequilibrium effects in
supersonic flows. Depending on the underlying physical models, numerical
simulations can be classified into three scales: macroscopic, microscopic,
and mesoscopic. Macroscopic-scale simulations rely on conservation equations
derived from the continuity assumption and near-equilibrium approximations,
including the Euler, NS, Burnett, and super-Burnett equations. The Euler
equations, based on equilibrium assumptions, cannot describe TNE effects.
The NS equations, based on near-equilibrium assumptions, incorporate linear
constitutive relations (first-order viscous stress and heat flux terms) to
represent TNE effects. However, in supersonic cross-regime flows, the
effects of deviating from thermodynamic equilibrium are highly complex,
profound, and multifaceted. Simply adding viscosity and heat conduction
terms to the hydrodynamic equations, particularly linear ones, is
insufficient to capture these effects accurately. As a result, the Euler and
NS equations are applicable only to studying supersonic flows in continuous
flow regimes ($Kn < 0.001$) where nonequilibrium effects are relatively
weak. For the slip-transition regime, the Burnett and super-Burnett
equations, derived from second- and third-order Chapman-Enskog expansions of
the velocity distribution function near thermodynamic equilibrium,
theoretically cover continuous, slip, and parts of the transitional regimes
with Knudsen numbers less than 1. However, these generalized macroscopic
equations still encounter significant challenges in practical applications,
including numerical instability at high Knudsen numbers, small wavelength
instability with grid refinement, high nonlinearity, difficulty in
satisfying entropy conditions, complex boundary conditions, intricate
programming, and low parallel computing efficiency. As a representative
microscopic method, molecular dynamics (MD) simulation is particularly
effective for studying nonequilibrium effects in supersonic flows at small
scales~\citep{liu2016molecular,liu2017molecular}. The strength of this
method lies in its ability to account for the microscopic structure and
dynamic behavior of molecules, thereby providing more accurate descriptions
and predictions. However, due to high computational costs and limited
spatial and temporal scales, MD is less suitable for large-scale,
long-duration simulations, often requiring integration with other methods
for a more comprehensive analysis.

As a mesoscopic approach bridging microscopic and macroscopic methods, the
Boltzmann equation links interactions across scales by describing the
spatiotemporal evolution of molecular velocity distributions, serving as a
key component in kinetic theory framework and addressing TNE effects across
the full spectrum of flow regimes. Kinetic modeling methods based on the
Boltzmann equation for studying multiscale nonequilibrium flows can be
broadly classified into the following four categories. (1) Kinetic
macroscopic modeling: This method involves deriving macroscopic hydrodynamic
equations (MHEs) through Chapman-Enskog multiscale analysis and Hermite
polynomial expansion of the Boltzmann equation, followed by numerical
solutions of these equations~\citep{Struchtrup-Book,zhao2014formulation}.
Consequently, the physical capabilities of these methods are confined to the
scope described by the MHEs. It is important to note that MHEs can be
divided into two categories: those involving only conserved moments, such as
the Burnett and super-Burnett equations, and extended MHEs that include the
evolution of both conserved and some non-conserved moments, such as Grad's
13-moment equations~\citep{2003-Struchtrup-R13}, the Eu method~%
\citep{eu1992kinetic}, and the nonlinear coupled constitutive relations
method~\citep{li2021unified,rana2021efficient,jiang2019computation}. (2)
Kinetic direct modeling: This approach directly simulates the (modified or
original) Boltzmann equation without deriving or solving MHEs, as in the
direct simulation Monte Carlo (DSMC) method~%
\citep{moss2005direct,schwartzentruber2015progress}. The DSMC method models gas dynamics in
nonequilibrium states by tracking the trajectories and collisions of gas
molecules, making it particularly suitable for high Knudsen number flows. In
situations where experimental data are difficult to obtain, DSMC simulation
results are often considered a reliable approximation of experimental
outcomes. However, due to constraints on grid cell size and time step (which
must be smaller than the mean free path and particle collision time,
respectively), the DSMC method becomes computationally intensive and
memory-demanding when simulating near-continuum flows.
(3) Direct solution of the Boltzmann equation: This category includes
methods such as the discrete velocity method~%
\citep{li2009gas,yang2016numerical}, Fourier spectral method, and fast
spectral method~\citep{wu2015fast}. From a physical perspective, directly
solving the original Boltzmann equation numerically preserves its physical
properties to the greatest extent. These methods, however, face challenges,
including high computational costs, significant memory requirements,
dependence on specific boundary conditions, and complexity in algorithm
implementation. (4) Kinetic discrete scheme method: Recent advances in
kinetic discrete scheme methods have enhanced the study of nonequilibrium
flows. Methods in this category include the gas kinetic scheme, unified gas
kinetic scheme, discrete unified gas kinetic scheme, unified gas kinetic
particle method, and unified gas kinetic wave particle method~%
\citep{xu2014direct,xu2010unified,guo2015discrete,zhu2019unified}, etc.
While these methods have proven effective, they also pose challenges,
including computational complexity, implementation difficulty, and, in some
cases, substantial demands on computational resources.

Despite significant progress in kinetic modeling based on the Boltzmann
equation, ``\emph{how to accurately describe}'' and ``\emph{how to visually
analyze}'' mesoscopic states and effects remain major challenges in the
numerical study of
supersonic flows. Compared with the macroscopic state, the mesoscopic states exhibit a higher degree of
discreteness and TNE. As a result, more variables are required to fully characterize the mesoscopic state.
In terms of description, most kinetic methods currently still adhere to
traditional macroscopic approaches, focusing primarily on a few physical
quantities in the NS model, such as slow variables, conserved quantities,
and a limited number of lower-order non-conserved quantities, lacking novel
perspectives. As the degree of discretization and TNE increases, the
complexity and uncertainty of system behaviors escalate, necessitating the
inclusion of additional physical quantities and broader perspectives,
particularly fast variables and higher-order non-conserved quantities across
various scales. On the analytical side, it is urgent to develop methods for
clearly and intuitively extracting and presenting information from complex
nonequilibrium physical fields, as this will determine the depth and
effectiveness of research. The technical key is how to achieve an intuitive
geometric correspondence in the description of states and behaviors in
complex systems. Commonly used measures of nonequilibrium in traditional
fluid dynamics, such as the Knudsen number, viscosity, heat conduction, and
macroscopic gradients, each describe the system's nonequilibrium state from
a specific perspective. However, these measures are highly condensed,
averaged, and coarse-grained, often obscuring specific details about the
nonequilibrium state that are crucial for in-depth analysis.

To tackle these challenges, we use the discrete Boltzmann
method/modeling/model (DBM) developed by our research group~%
\citep{xu2012lattice,xu-Book,xu2024advances}. DBM is essentially a
coarse-grained modeling and analysis method that combines kinetic theory
with mean field theory. Historically, DBM evolved from the physical modeling
branch of the lattice Boltzmann method (LBM)~%
\citep{succi2018lattice,guo2013lattice,wei2022small}, selectively retaining, discarding,
and enhancing various aspects. Unlike traditional fluid modeling, DBM does
not rely on the continuity assumption or near-equilibrium
approximations.  It
no longer uses the standard ``lattice gas'' imagery of the LBM,
nor does it combine with specific discretization schemes~\citep{wang2020simple,wang2022novel}.
Instead, it introduces kinetic schemes for detecting, presenting, describing, and analyzing
nonequilibrium states and effects based on phase space theory.
As the degree of discretization and nonequilibrium increases, DBM
distinguishes itself from traditional modeling methods and other kinetic
approaches by incorporating more kinetic modes in physical modeling and
utilizing higher-order non-conserved quantities, particularly TNE
quantities, in physical analysis to describe system states and behaviors.
Consequently, from a modeling perspective, DBM is a direct kinetic modeling
approach, while from the perspective of complex physical field analysis, it
serves as a kinetic analysis method.

In recent years, DBM has made significant advances in the study of nonequilibrium flows and transport mechanisms, with applications in high-speed compressible flows~\citep{2013-Gan-EPL,gan2018discrete,bao2022study,Qiu-POF-2020,Qiu-PRE-2021,qiu2024mesoscopic}, multiphase flows~\citep{2015-Gan-SM-PS,gan2022discrete,2019-Zhang-SM-PS}, fluid instabilities~\citep{gan2019nonequilibrium,lai2016nonequilibrium,chen2024surface,chen2024kinetics}, micro- and nanoscale flows~\citep{zhang2023lagrangian}, chemically reactive flows~\citep{lin2017multi,ji2022three}, and plasma kinetics~\citep{liu2023discrete,song2024plasma}.
The gradual advancement of DBM research from the ``shallow water" regime (small Kn number) to the ``deep water" regime (increasing Kn number) is an ongoing effort we have been making in recent years~\citep{gan2018discrete,gan2022discrete,zhang2019discrete,zhang2022discrete,song2023,
song2024plasma,shan2023nonequilibrium}.
For example, DBM has successfully reproduced and kinetically  clarified heat conduction behavior in square cavity flow~\citep{zhang2019discrete} and following strong shock wave~\citep{shan-}, phenomena that appear to violate Fourier's law.
DBM studies revealthat nonequilibrium effects near mesoscale structures, such as material, shock wave, and liquid-vapor interfaces, exhibit strong nonlinearity. The relative strength of higher- and lower-order nonequilibrium effects is significant, with higher-order nonequilibrium influencing lower-order effects through a complex feedback relationship~\citep{gan2018discrete,gan2022discrete}.
In the steady-state DBM, which focuses on steady-state behavior, the distribution function transitions from accounting for only partial Kn number effects in the time-dependent scenarios to including all Kn number effects~\citep{zhang2023lagrangian}.
It should be noted that the mesoscale temporal behavior, sandwiched between macroscopic continuous modeling and molecular dynamics, remains poorly understood.

In supersonic flows, as Mach numbers increase, compressibility intensifies,
rarefied gas effects become more pronounced, complex interfaces arise, and
small-scale structures emerge, the types, magnitudes, and durations of
nonequilibrium driving forces in the system multiply. This necessitates
reassessing the validity of existing DBM models and reconsidering current
research perspectives and analytical approaches.
Therefore, this paper seeks to further explore the following
questions:
(1) How to accurately, efficiently, and conveniently identify and incorporate the
necessary kinetic modes under these extreme conditions to precisely describe
strong nonequilibrium effects?
(2) How to derive theoretical expressions for
higher-order nonequilibrium effects influenced by multiple physical fields?
(3) How to develop cross-scale DBM suitable for complex fluid systems with
higher discretization and stronger nonequilibrium effects?
(4) How to design
coupling and switching criteria for multiscale DBM to achieve adaptive and
efficient simulations of multiscale problems?
(5) How to adjust perspectives on
nonequilibrium effect descriptions, enhance analysis methods for
nonequilibrium behaviors, and expand the depth and breadth of research into
nonequilibrium effects?
(6) Finally, how to use DBM simulations and phase-space
analysis method to clarify the nonequilibrium characteristics and evolution
mechanisms of mesoscale structures in supersonic flows, as well as their
impact on flow, heat transfer, and entropy production mechanisms?

\section{Physical modeling of thermodynamic nonequilibrium effects}

\label{modeling}

The DBM encompasses two critical components: modeling TNE effects and
extracting, presenting, and analyzing these effects.
The accuracy in describing TNE effects
directly determines the model's precision, as TNE effects are fundamental
characteristics of multiscale systems, which are intrinsically tied to the
constitutive relations of the fluid system and their evolution.

TNE effects consist of two primary aspects: spatial nonequilibrium (discrete
effects) and temporal nonequilibrium (related to relaxation time). TNE
effects exhibit a dual nature. On one hand, they represent a hallmark of
nonequilibrium characteristics, reflecting the complexity of the system. On
the other hand, they act as the driving force for system evolution. TNE
effects and mesoscale structures are deeply coupled and mutually
influential. Their interactions can be classified into: (1) driving and
formation, (2) enhancement and feedback, (3) stabilization and
self-organization, (4) competition and selection, (5) feedback and
regulation. In this section, we discuss how to realize multiscale modeling
and analyzing TNE effects via DBM.

\subsection{Thermodynamic nonequilibrium effects: view from macroscopic
hydrodynamic equations and DBM}

By performing the Chapman-Enskog multi-scale analysis on the Boltzmann-BGK
equation, macroscopic hydrodynamic equations (MHEs) can be derived. To
simplify the complex collision term in the original Boltzmann equation and
adjust the Prandtl number, a linearized collision term with Shakhov model is
used~\citep{shakhov1968generalization}
\begin{equation}
\frac{{\partial f}}{{\partial t}}+\mathbf{v} \cdot \bm{\nabla }f=-\frac{1}{%
\tau }(f-{f^{S}}),  \label{Boltzmann-BGK}
\end{equation}
where $f$, $\mathbf{v}$, and $\tau$ denote the distribution function,
particle velocity, and relaxation time, respectively.
In fact, the BGK operator employed in nonequilibrium flow research is modified using mean-field theory~\citep{gan2022discrete,xu2024advances}.
The Shakhov
distribution function is given by ${f^{S}}={f^{eq}}+{f^{eq}}\{(1-\Pr )%
\mathbf{v}^*\cdot \mathbf{q}[\frac{\mathbf{v}^{\ast }{{^{2}}+\bm{\eta }{^{2}}%
}}{{RT}}-(D+n+2)]/[(D+n+2)PRT]\}$, where $\Pr$, $\mathbf{q}$, $R$, $T$, $%
P(=\rho R T)$, and $D$ represent the Prandtl number, heat flux, gas
constant, temperature, pressure, and spatial dimension, respectively. $%
\mathbf{v}^*=\mathbf{v}-\mathbf{u}$ is the thermo-fluctuation velocity and $%
\mathbf{u}$ the fluid velocity. $\bm{\eta}^2=\sum_{j=1}^{n}\bm{\eta
}{_{j}^{2}}$, with $\bm{\eta}_j$ being a free parameter introduced to
account for the $n$ extra degrees of freedom corresponding to molecular
rotation and/or vibration. Finally, ${f^{eq}}=\rho {(\frac{1}{{2\pi RT}}%
)^{(D+n)/2}}\exp [-\frac{\mathbf{v}^{\ast }{{^{2}}}}{{2RT}}-\frac{\bm{\eta
}{{^{2}}}}{{2RT}}]$ represents the Maxwell equilibrium distribution
function.
The Shakhov model adds a nonequilibrium term into the distribution
function to adjust energy transport, thereby partially addressing the BGK
model's limitation of a fixed Prandtl number ($\Pr = 1$). This enhancement
allows for a more accurate depiction of system behavior under nonequilibrium
conditions. While utilizing ${f^{S}}$ instead of ${f^{eq}}$ increases the
complexity of the collision term, this added complexity is essential for
achieving greater physical accuracy. Consequently, practical applications
should carefully balance computational complexity and the desired level of
accuracy in the results.

Applying Chapman-Enskog expansion to both sides of equation %
\eqref{Boltzmann-BGK} yields the following extended MHEs:
\begin{equation}
\left\{
\begin{array}{l}
{\partial }_{t}{\rho }+\bm{\nabla }\cdot ({\rho }\mathbf{u})=0, \\
{\partial }_{t}({\rho }\mathbf{u})+\bm{\nabla }\cdot ({\rho }\mathbf{uu}+P%
\mathbf{I}+\bm{\Delta }{_{2}^{\ast }})=0, \\
{\partial }_{t}({\rho }e)+\bm{\nabla }\cdot \lbrack ({\rho }e+P)\mathbf{u}+%
\bm{\Delta }{_{2}^{\ast }}\cdot \mathbf{u}+\bm{\Delta }_{3,1}^{\ast })=0.%
\end{array}%
\right.  \label{Burnett}
\end{equation}
Here, $e=c_{v}T+u^{2}/2$ the specific total energy with $c_{v}=(n+2)R/2$ the specific heat at constant volume, $\bm{\Delta }_{2}^{\ast }$ and $\bm{\Delta }_{3,1}^{\ast }$ correspond
to the generalized viscous stress and heat flux, respectively. These
extended MHEs incorporate higher-order constitutive corrections, offering a
more comprehensive representation of TNE phenomena and their impact on fluid
dynamics.

Equation \eqref{Burnett} demonstrates that MHEs describe TNE effects
exclusively and implicitly through constitutive relations. In contrast, DBM
captures TNE effects directly and explicitly via higher-order non-conserved
moments of $(f-f^{(0)})$:
\begin{equation}
\bm{\Delta}_{m,n}^{\ast }=\mathbf{M}_{m,n}^{\ast }(f-f^{(0)})=\int_{-\infty
}^{\infty }\int_{-\infty }^{\infty }(\frac{1}{2})^{1-\delta _{m,n}}(f-f^{(0)})%
{\underbrace{{\mathbf{v}}^{\ast }{\mathbf{v}}^{\ast }\cdots {\mathbf{v}}%
^{\ast }}_{n}(\mathbf{v}}^{\ast 2}+{\mathbf{\eta }{^{2}})}^{\frac{m-n}{2}}d%
\mathbf{v}d\mathbf{\eta }{_{i}}\text{,}
\end{equation}%
\label{TNE1} where $\bm{\Delta }_{m,n}^{\ast }$ represents $m$-th order
tensors contracted to $n$-th order tensors, with $\delta _{m,n}$ as the
Kronecker delta function, and $f^{(0)}=f^{eq}$. When $m=n$, $\bm{\Delta }%
_{m,n}^{\ast }$ reduces to $\bm{\Delta }_{m}^{\ast }$. Replacing $\mathbf{v}%
^{\ast }$ in equation \eqref{TNE1} with $\mathbf{v}$ converts the central
moment $\bm{\Delta }_{m,n}^{\ast }$ into the non-central moment $\bm{\Delta}%
_{m,n}$, which describe the combined effects of HNE and TNE, typically
referred to as thermo-hydrodynamic non-equilibrium (THNE) effects.
Additionally, $\bm{\Delta}_{m,n}-\bm{\Delta}_{m,n}^{\ast }$ represents HNE
effects, offering a valuable supplement to the description of nonequilibrium
effects based on macroscopic gradients in traditional fluid mechanics. TNE
measures ($\bm{\Delta }_{2}^{\ast }$ and $\bm{\Delta }_{3,1}^{\ast }$) in
MHEs are typical ones as defined by equation \eqref{TNE1}, also known as
non-organized moment flux (NOMF) and non-organized energy flux (NOEF).

Moreover, it can be observed that MHEs describe only the evolution of the
three conserved moments $(\rho ,\rho \mathbf{u},\rho e)=(\mathbf{M}_{0}(f),%
\mathbf{M}_{1}(f),\mathbf{M}_{2,0}(f))$. Higher-order non-conserved moments
of $f$ are not included in traditional fluid dynamics. This has two
implications: it simplifies macroscopic hydrodynamic descriptions but also
creates challenges in accurately describing discontinuous or nonequilibrium
flows. In contrast, recovering the corresponding level of MHEs is just one
of the DBM's physical functions. Corresponding to the physical functions of
DBM are the extended hydrodynamic equations (EHEs), which, in addition to
equation \eqref{Burnett}, also include evolution equations for some of the
most closely related non-conserved moments, such as, $\mathbf{M}_{2}(f),%
\mathbf{M}_{3,1}(f),\mathbf{M}_{3}(f)$ and $\mathbf{M}_{4,2}(f)$, etc. The
necessity of these extended parts (i.e., the evolution equations of the
relevant non-conserved moments) increases rapidly with the degree of
discretization and nonequilibrium, as they describe the evolution of
nonlinear constitutive relations (equations \eqref{E1} and \eqref{E2}) and
so on:
\begin{equation}
{\partial }_{t}\bm{\Delta }_{2}^{\ast }+{\partial }_{t}\mathbf{M}_{2}^{\ast
}(f^{(0)})+\bm{\nabla}\cdot \mathbf{M}_{3}^{\ast }(f^{(0)})+\mathbf{M}%
_{2}^{\ast }(f^{(0)})\mathbf{u}+\bm{\Delta }_{3}^{\ast }+\bm{\Delta }%
_{2}^{\ast }\mathbf{u}]+\frac{1}{\tau }\bm{\Delta }_{2}^{\ast }=0,
\label{E1}
\end{equation}%
\begin{equation}
{\partial }_{t}\bm{\Delta }_{3,1}^{\ast }+{\partial }_{t}\mathbf{M}%
_{3,1}^{\ast }(f^{(0)})+\bm{\nabla}\cdot \mathbf{M}_{4,2}^{\ast }(f^{(0)})+%
\mathbf{M}_{3,1}^{\ast }(f^{(0)})\mathbf{u}+\bm{\Delta }_{4,2}^{\ast }+%
\bm{\Delta }_{3,1}^{\ast }\mathbf{u}]+\frac{1}{\tau }\bm{\Delta }_{3,1}^{\ast
}=0\text{.}  \label{E2}
\end{equation}

\subsection{Derivation of thermodynamic nonequilibrium effects}

The TNE measures, $\bm{\Delta}_{m,n}^{\ast }$, are primarily unknown. To
derive their explicit expressions, we perform the Chapman-Enskog expansion
on both sides of equation \eqref{Boltzmann-BGK} by introducing the following
expansions:
\begin{equation}
\left\{
\begin{array}{l}
f=f^{(0)}+f^{(1)}+f^{(2)}+\cdots, \\
f^{S}=f^{(0)}+f^{S(1)}+f^{S(2)}+\cdots, \\
\frac{\partial }{\partial t}=\frac{\partial }{\partial {{t}_{1}}}+\frac{%
\partial }{\partial {{t}_{2}}}+\cdots, \\
\bm{\nabla}=\bm{\nabla}_{1},%
\end{array}%
\right.
\end{equation}
where $f^{(j)}$, $f^{S(j)}$, $\partial _{t_{j}}$, $\bm{\nabla}_{j}$
represent the $j$-th order terms in the Knudsen number. By comparing the
coefficients at each order, we obtain the first- and second-order deviations
of the distribution function:
\begin{eqnarray}
f^{(1)} &=&-\tau (\frac{\partial f^{(0)}}{\partial {{t}_{1}}}+\mathbf{v}%
\cdot {{\bm{\nabla} }_{1}}f^{(0)})+f^{S(1)}, \\
f^{(2)} &=&-\tau (\frac{\partial f^{(0)}}{\partial {{t}_{2}}}+\frac{\partial
f^{(1)}}{\partial {{t}_{1}}}+\mathbf{v}\cdot {{\bm{\nabla} }_{1}}%
f^{(1)})+f^{S(2)}.
\end{eqnarray}
Next, we derive the TNE expressions by using transformation relations for
temporal and spatial derivatives at different scales, $\partial {t_{j}}(.)$
and $\bm{\nabla}_{j}(.)$, along with the kinetic moments of $f^{S(j)}$,
given by $\mathbf{M}_{2}({{f}^{S(n)}})=0$ and $\mathbf{M}_{3,1}({{f}^{S(n)}}%
)=(1-\Pr )\mathbf{q}^{(n)}$.

After some straightforward but tedious algebraic manipulations, we obtain
the following relations between thermodynamic forces and fluxes:
\begin{equation}
\bm{\Delta }_{2}^{\ast (1)}=-\mu [\bm{\nabla} \mathbf{u}+{(\bm{ \nabla}
\mathbf{u})}^{T}-\frac{2}{n+2}{\mathbf{I} \bm{\nabla} \cdot \mathbf{u}}],
\label{vis-1st}
\end{equation}
\begin{equation}
\bm{\Delta }_{3,1}^{\ast (1)}=-\kappa {\bm{\nabla }}T,  \label{heat-1st}
\end{equation}
where $\mu =P \tau $ is the viscosity coefficient, and $\kappa = c_p \mu/\Pr
$ is the thermal conductivity, with $c_{p}=(n+4)R/2$ the specific heat at
constant pressure. The expressions for the second-order constitutive
relations, $\bm{\Delta }_{2}^{\ast(2)}$ and $\bm{\Delta }_{3,1}^{\ast(2)}$,
can be found in Appendix \ref{appendix1}. Consequently, the constitutive
relations at the Burnett level are given by $\bm{\Delta }_{2}^{\ast }=%
\bm{\Delta}_{2}^{\ast (1)}+\bm{\Delta }_{2}^{\ast (2)}$ and $\bm{\Delta }%
_{3,1}^{\ast }=\bm{\Delta}_{3,1}^{\ast (1)}+\bm{\Delta }_{3,1}^{\ast (2)}$,
which are expected to significantly enhance macroscopic modeling. Similar
derivations can be used to obtain counterparts and other TNE measures at and
beyond the super-Burnett level.

Analysis of the above analytical expressions reveals that:

(i) TNE intensity depends on macroscopic quantities, their gradients,
relaxation time, additional degrees of freedom, and the Prandtl number.
Here, relaxation time characterizes temporal nonequilibrium, whereas
gradients of macroscopic quantities describe spatial nonequilibrium or
discreteness. As TNE intensity increases, the expressions incorporate more
physical quantities and their gradients. This reflects a broader range of
nonequilibrium driving forces and highlights the increasing complexity of
these effects. For a detailed analysis, please refer to Appendix \ref%
{appendix1}.

(ii) As TNE intensity increases, the linear constitutive relations fail, a
Newtonian fluid will transition into a non-Newtonian fluid. When
higher-order TNE effects become non-negligible, the system exhibits
characteristics akin to non-Newtonian fluids, such as shear thinning, shear
thickening, thixotropy, viscoelasticity, and antithixotropy. These behaviors
arise from different types of feedback from higher-order TNE effects on
constitutive relations.

(iii) The Shakhov-BGK model adjusts the system's Prandtl number by
introducing a heat flux term. However, this approach is only suitable for
cases with weak TNE effects. As TNE intensity increases, the system's
Prandtl number is no longer a fixed value depending solely on material
properties. Instead, it becomes a variable that also depends on the local
state of the flow field. At this stage, a generalized expression can be
defined as: ${\Pr}^* = \frac{\mu^* C_p^*}{\kappa^*}$, where $\mu^*$, $C_p^*$%
, and $\kappa^*$ represent the effective dynamic viscosity, specific heat
capacity, and thermal conductivity, respectively. These quantities can be
expressed as corrections to their classical values: $\mu^* = \mu + \Delta
\mu (\partial_{\alpha} u_{\beta}, \partial_{\alpha} \rho, \partial_{\alpha}
T, \dots)$, $C_p^* = C_p + \Delta C_p (\partial_{\alpha} u_{\beta},
\partial_{\alpha} \rho, \partial_{\alpha} T, \dots)$, and $\kappa^* = \kappa
+ \Delta \kappa (\partial_{\alpha} u_{\beta}, \partial_{\alpha} \rho,
\partial_{\alpha} T, \dots)$, with $\Delta \mu$, $\Delta C_p$, and $\Delta
\kappa$ the corrections to the dynamic viscosity, specific heat capacity,
and thermal conductivity due to higher-order TNE effects, respectively.
Additionally, specific expressions may also need to consider fluid
properties, boundary conditions, and external field influences.

\subsection{Determination of the most essential moments}

The distribution functions $f^{eq}$ and $f^S$ possess an infinite number of
kinetic moments, each describing different aspects of the fluid system's
kinetic properties. In coarse-grained physical modeling, selecting kinetic
moments corresponds to choosing the physical properties that the model aims
to describe, which is a fundamental task in this approach~%
\citep{gan2022discrete}. To derive MHEs at the Burnett level from the
Boltzmann-BGK equation, nine independent kinetic moments of $f^{eq}$ are
used: ${\mathbf{M}}_0$, ${\mathbf{M}}_1$, ${\mathbf{M}}_{2,0}$, ${\mathbf{M}}%
_2$, ${\mathbf{M}}_{3,1}$, ${\mathbf{M}}_3$, ${\mathbf{M}}_{4,2}$, ${\mathbf{M}%
}_4$, and ${\mathbf{M}}_{5,3}$. Additionally, the following seven independent
kinetic moments of $f^S$ are used: ${\mathbf{M}}_0^S$, ${\mathbf{M}}_1^S$, ${%
\mathbf{M}}_{2,0}^S$, ${\mathbf{M}}_2^S$, ${\mathbf{M}}_{3,1}^S$, ${\mathbf{M}%
}_3^S$, and ${\mathbf{M}}_{4,2}^S$. Among these moments, the first three are
conserved moments, while the others are non-conserved. The moments ${\mathbf{%
M}}_3$, ${\mathbf{M}}_{4,2}$, ${\mathbf{M}}_2^S$, and ${\mathbf{M}}_{3,1}^S$
describe first-order TNE effects, while ${\mathbf{M}}_4$, ${\mathbf{M}}_{5,3}$%
, ${\mathbf{M}}_3^S$, and ${\mathbf{M}}_{4,2}^S$ correspond to second-order
TNE effects.

\subsection{Discretizations of velocity space and equilibrium distribution
function}

The most essential moments are the invariants to be preserved in DBM
coarse-grained physical modeling, i.e., these kinetic moments should remain
consistent when transforming from integral form to summation form
\begin{equation}
\bm{\Phi}_{m,n}=\int_{-\infty }^{\infty }\int_{-\infty }^{\infty }f^{(0)}%
\bm{\Psi}(\mathbf{v},\mathbf{\eta })\,d{\mathbf{v}}d{\mathbf{\eta }}%
=\sum\limits_{i}f_{i}^{(0)}\bm{\Psi}(\mathbf{v}_{i},\mathbf{\eta }_{i}),
\label{Mfeq}
\end{equation}%
\begin{equation}
\bm{\Phi}_{m,n}^{S}=\int_{-\infty }^{\infty }\int_{-\infty }^{\infty }f^{S}%
\bm{\Psi}^{\prime }(\mathbf{v},\mathbf{\eta })\,d{\mathbf{v}}d{\mathbf{\eta }%
}=\sum\limits_{i}f_{i}^{S}\bm{\Psi}^{\prime }(\mathbf{v}_{i},\mathbf{\eta }%
_{i}),  \label{Mfs}
\end{equation}%
where $\bm{\Phi}_{m,n}=[{\mathbf{M}}_{0},{\mathbf{M}}_{1},{\mathbf{M}}_{2,0},{%
\mathbf{M}}_{2},{\mathbf{M}}_{3,1},{\mathbf{M}}_{3},{\mathbf{M}}_{4,2},{%
\mathbf{M}}_{4},{\mathbf{M}}_{5,3}]$ and $\bm{\Phi}_{m,n}^{S}=[{\mathbf{M}}%
_{0}^{S},{\mathbf{M}}_{1}^{S},{\mathbf{M}}_{2,0}^{S}, \newline
{\mathbf{M}}_{2}^{S},{\mathbf{M}}_{3,1}^{S},{\mathbf{M}}_{3}^{S},{\mathbf{M}}%
_{4,2}^{S}]$. Correspondingly, $\bm{\Psi}(\mathbf{v},\mathbf{\eta })=[1,%
\mathbf{v},\frac{1}{2}(v^{2}+{\eta }^{2}),...,\frac{1}{2}(v^{2}+{\eta }^{2})%
\mathbf{vvv}]$, and $\bm{\Psi}^{\prime }(\mathbf{v},\mathbf{\eta })=[1,%
\mathbf{v},\frac{1}{2}(v^{2}+{\eta }^{2}),...,\frac{1}{2}(v^{2}+{\eta }^{2})%
\mathbf{vv}]$. By incorporating the additionally required kinetic moments
into $\bm{\Phi}_{m,n}$ and $\bm{\Phi}_{m,n}^{S}$, one can construct a DBM with
the desired order of accuracy and perform simulations without needing the
exact form of the extended MHEs. Compared to deriving and solving
higher-order extended MHEs, the complexity of constructing and computing DBM
only increases slightly. For example, to achieve third-order accuracy in
viscous stress and heat flux, incorporating $\mathbf{M}_{5}$ and $\mathbf{M}%
_{64}$ into $\bm{\Phi}_{m,n}$, and $\mathbf{M}_{4}^{S}$ and $\mathbf{M}%
_{5,3}^{S}$ into $\bm{\Phi}_{m,n}^{S}$, is sufficient. This makes DBM
particularly suitable for multiscale fluid simulations.

Equations \eqref{Mfeq} and \eqref{Mfs} can be rexpressed in matrix form as
follows:
\begin{equation}
\bm{\Phi}_{m,n} = \bm{\Psi} \mathbf{f}^{(0)}\text{,}  \label{f0}
\end{equation}%
\begin{equation}
\bm{\Phi}_{m,n}^S = \bm{\Psi}^{\prime }\mathbf{f}^{S}\text{.}  \label{fs}
\end{equation}
Consequently, the discrete equilibrium distribution function (DEDF) vectors $%
\mathbf{f}^{(0)}$ and $\mathbf{f}^{S}$ can be computed as follows~%
\citep{2013-Gan-EPL},
\begin{equation}
\mathbf{f}^{(0)}=\bm{\Psi}{^{-1}} \bm{\Phi}_{m,n}\text{,}  \label{f0}
\end{equation}%
\begin{equation}
\mathbf{f}^{S}=\bm{\Psi}^{\prime }{^{-1}} \bm{\Phi}_{m,n}^S\text{,}
\label{f0}
\end{equation}%
where $\bm{\Psi}^{-1}$ and $\bm{\Psi}^{\prime -1}$ represent the inverses of
$\bm{\Psi}$ and $\bm{\Psi}^{\prime }$, respectively. To improve the physical
accuracy and computational stability of the model, we require that $\mathbf{f%
}^{(0)}$ further satisfy the moment relation $\mathbf{M}_{40}$, and $\mathbf{%
f}^{S}$ satisfy the moment relations $\mathbf{M}^S_{4}$, $\mathbf{M}^S_{5,3}$%
, and $\mathbf{M}^S_{40}$. Thus, $\bm{\Phi}%
_{m,n}=(M_{0},M_{1x},M_{1y},M_{2,0},...,M_{5,3yyy},M_{40})^{T} $, $\bm{\Phi}%
_{m,n}^S=(M_{0}^S,M_{1x}^S,M_{1y}^S,M_{2,0}^S,...,M_{5,3yyy}^S,M_{40}^S)^{T}$,
$\mathbf{f}^{(0)}=(f_{1}^{(0)},f_{2}^{(0)},...,f_{26}^{(0)})^{T}$, $\mathbf{f%
}^{S}=(f_{1}^{S},f_{2}^{S},...,f_{26}^{S})^{T}$, where each vector has 26
components in the two-dimensional case. The matrix $\bm{\Psi} = \bm{\Psi}%
^{\prime }= (\mathbf{C}_{1}, \mathbf{C}_{2}, \cdots, \mathbf{C}_{26})$ is a $%
26 \times 26$ matrix linking the DEDF and the kinetic moments, with $\mathbf{%
C}_{i} = [1, v_{ix}, v_{iy}, \ldots, \frac{1}{2}(v_{ix}^2 + v_{iy}^2 +
\eta_i^2)v_{iy}^3,\frac{1}{2}(v_{ix}^2 + v_{iy}^2 + \eta_i^2)^2]^{T}$.

DBM uses a finite number of particle velocities to replace the continuous
velocity space. After discretizing, the discrete form of equation %
\eqref{Boltzmann-BGK} is:
\begin{equation}
\frac{{\partial f_i}}{{\partial t}}+\mathbf{v}_i \cdot \bm{\nabla }f_i=-%
\frac{1}{\tau }(f_i-{f_i^{S}}).  \label{Boltzmann-BGK2}
\end{equation}
The evolution equation \eqref{Boltzmann-BGK2} for $f_i$ and the constraint
equations \eqref{f0}-\eqref{fs} for the discrete velocities $(\mathbf{v}_i, %
\bm{\eta}_i)$ together form the model equation of the DBM. In principle, the
choice of discrete velocities is arbitrary, as long as the matrices $%
\bm{\Psi}$ and $\bm{\Psi}^{\prime }$ are of full rank. In practical
simulations, the discrete velocity stencil typically aims to improve
numerical stability by maintaining high geometric symmetry while ensuring
physical symmetry (i.e., satisfying the corresponding moment relations).
Since the number of discrete velocities equals the number of independent
moment relations, this discretization method is physically the most
computationally efficient. Strictly speaking, discretizing the continuous
space inevitably breaks physical symmetry, leading to inaccurate
descriptions of system properties. Therefore, coarse-grained physical
modeling inherently involves trade-offs: the system characteristics under
study must be strictly preserved during discretization, while
characteristics beyond the scope of the study can be selectively retained or
sacrificed. Among the many options that meet the research needs, the
simplest or most economical is often the preferred choice.

\subsection{Extraction and exhibition of thermodynamic nonequilibrium effects%
}

\begin{figure}
\centering{%
\epsfig{file=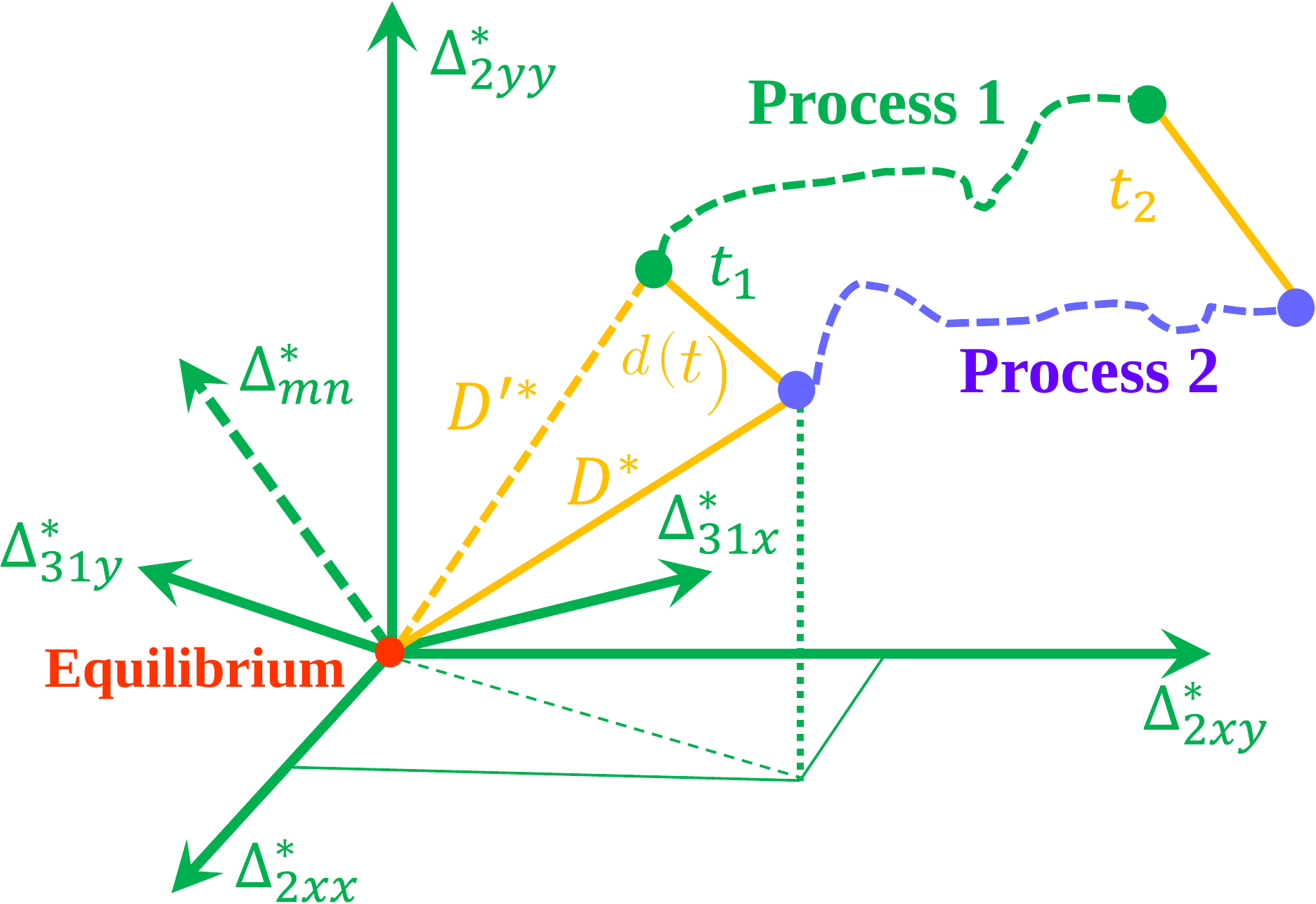,
			width=0.55\textwidth}}
\caption{Schematic diagram of the phase space spanned by the independent
components of the nonequilibrium characteristic quantity $\bm{\Delta}_{m,n}$.}
\label{ps}
\end{figure}

DBM is not only a multiscale modeling method for nonequilibrium flows but
also an analysis tool for complex physical fields. In other words, it
handles both pre-simulation physical modeling and post-simulation data
analysis. To intuitively display complex field information, DBM has
developed a scheme based on phase space for detecting, presenting,
describing, and analyzing nonequilibrium states and effects. Specifically,
DBM employs the independent components of non-conservative moments of $%
(f-f^{eq})$, $\bm{\Delta}_{m,n}^{\ast}$, to construct a high-dimensional
phase space that describes the system's nonequilibrium states and effects,
as illustrated in figure \ref{ps}. In this phase space, the origin
represents the equilibrium state, and any other point indicates a
nonequilibrium state. The intercepts of this nonequilibrium state on each
axis describe the degree and manner of the system's deviation from
equilibrium from their respective perspectives. The distance from a
nonequilibrium point to the origin roughly measures the degree of
nonequilibrium, or the total nonequilibrium intensity,
\begin{equation}
D_T^{\ast } = \sqrt{\sum\nolimits_{m,n}{\bm{\Delta}_{m,n}^{\ast 2}}}.
\end{equation}
Accordingly, the dimensionless total nonequilibrium intensity is defined by
\begin{equation}
\tilde D_T^{\ast } = \sqrt{\sum\nolimits_{m,n}({\bm{\Delta}_{m,n}^{\ast}/T^{%
\frac{m}{2}})^2}}.
\end{equation}
The reciprocal of the distance between two points defines the similarity
between two nonequilibrium states. Further, introducing a nonequilibrium
intensity vector, $\mathbf{S}_{\text{TNE}}$, enables a multi-perspective
description of the nonequilibrium state
\begin{equation}
\mathbf{S}_{\text{TNE}}=\left\{
\begin{array}{c}
\Delta _{2xx}^{\ast },\Delta _{2xy}^{\ast },\Delta _{2yy}^{\ast },\Delta
_{3,1x}^{\ast },\Delta_{3,1y}^{\ast},...,\Delta _{5,3xyy}^{\ast },\Delta
_{5,3yyy}^{\ast }, \\
|\bm{\Delta}_{2}|,|\bm{\Delta}_{3,1}|,|\bm{\Delta}_{3}|,|\bm{\Delta}_{4,2}|,|%
\bm{\Delta}_{4}|, |\bm{\Delta}_{5,3}|,D_{T}^{\ast },\tilde{D}_{T}^{\ast }%
\end{array}%
\right\} ,
\end{equation}
where the first line specifies the components of $\bm{\Delta}_{m,n}$, the
second line quantifies the intensity of each TNE measure and the overall
nonequilibrium strength. The phase space method, based on the
non-conservative moments of $(f-f^{eq})$, provides clear, intuitive
geometric representations of complex nonequilibrium states and behaviors,
simplifying the understanding of complex systems and enabling detailed,
quantifiable research of previously elusive information.

\section{ Numerical results and analysis}

\label{Numerical simulations}

In this section, we first develop the first-order and second-order DBMs.
Then, we validate the second-order model's ability to describe large-scale
structures through a series of benchmarks, ranging from subsonic to
hypersonic regimes. Next, we assess the model's effectiveness in capturing
nonequilibrium effects across various orders. Finally, we investigate both
HNE and TNE characteristics, along with
entropy production mechanisms, during the regular reflection of shock waves.

\begin{figure}
\centering{%
\epsfig{file=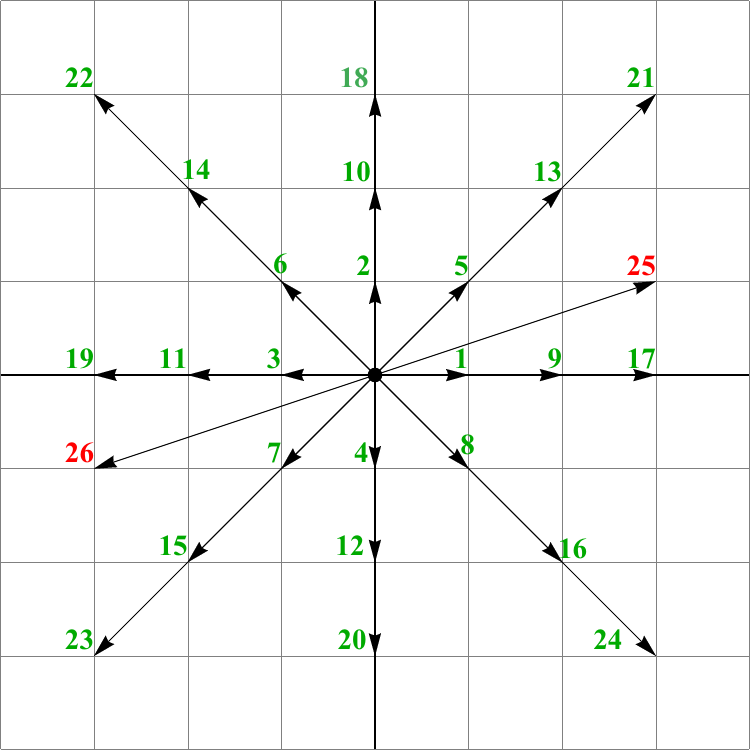,
			width=0.55\textwidth}}
\caption{Sketch of the discrete velocity stencil. The first-order DBM uses
the discrete velocities $1-16$, and the second-order DBM uses the discrete
velocities $1-26$.}
\label{D2V26}
\end{figure}

To perform numerical simulations, it is necessary to design a set of
discrete velocities to discretize the velocity space and ensure the
existence of $\bm{\Psi}{^{-1}}$ and $\bm{\Psi}^{\prime }{^{-1}}$. Figure \ref%
{D2V26} illustrates the discrete velocity stencil, with the first-order DBM
using the first 16 discrete velocities and the second-order DBM utilizing
all 26 velocities:
\begin{equation}
(v_{ix},v_{iy})=\left\{
\begin{array}{cc}
\text{cyc}:c(\pm 1,0) & \text{for}\quad 1\leq i\leq 4 \\
c(\pm 1,\pm 1) & \text{for}\quad 5\leq i\leq 8 \\
\text{cyc}:2c(\pm 1,0) & \ \text{for}\quad 9\leq i\leq 12 \\
2c(\pm 1,\pm 1) & \ \quad \text{for}\quad 13\leq i\leq 16 \\
\text{cyc}:3c(\pm 1,0) & \quad \ \text{for}\quad 17\leq i\leq 20 \\
3c(\pm 1,\pm 1) & \quad \ \text{for}\quad 21\leq i\leq 24 \\
c(a,b),-c(a,b) & \quad \ \text{for}\quad 25\leq i\leq 26%
\end{array}%
\right. ,  \label{DVM2}
\end{equation}
where ``cyc" indicates a cyclic permutation, $c$ is an adjustable parameter
controlling the size of discrete velocity, and $a$, $b$ are free parameters.
As a special case, $a=3$ and $b=1$ in figure \ref{D2V26}. For the
first-order DBM, when $1 \leq i \leq 4$, $\eta_i = \eta_0$, and for all
other $i$, $\eta_i = 0$. For the second-order DBM, when $1 \leq i \leq 4$, $%
\eta_i = i \eta_0 $, and when $13 \leq i \leq 20$, $\eta_i = \eta_0$;
otherwise, $\eta_i = 0$.

It is important to note that the physical model presented in Section \ref%
{modeling}, derived from coarse-grained modeling, does not include any
specific discretization scheme. The choice of discrete velocities is a key
technical aspect of DBM simulations. The discrete velocity set in equation %
\ref{DVM2} is not a universal or optimal standard for all cases, but rather
a method that meets the current research requirements. The selection of
optimal discrete velocities is a complex problem involving the coupling of
phase space discretization with temporal and spatial discretizations. In
recent work, we aim to explore this issue in-depth via von Neumann stability
analysis.

To improve numerical stability, efficiency, and accuracy in capturing
complex multiscale structures, the third-order implicit-explicit Runge-Kutta
finite difference scheme [94] is employed for time discretization; and
depending on stability requirements, either the second-order
non-oscillatory, parameter-free dissipative finite difference (FD) scheme or
the fifth-order weighted essentially non-oscillatory FD scheme is applied
for spatial discretization.

All physical quantities in the simulations are normalized using reference
density $\rho_0$, reference temperature $T_0$, and reference length $L_0$:
\begin{equation}
\left\{
\begin{array}{l}
\hat{\rho}=\frac{\rho }{{\rho _{0}}},\text{ \ }{\hat{r}_{\alpha }}=\frac{{%
r_{\alpha }}}{{L_{0}}},\text{ \ }\hat{T}=\frac{T}{{T_{0}}},\text{ \ }(\hat{t}%
,\hat{\tau})=\frac{{(t,}\text{ }{\tau )}}{{{L_{0}}/\sqrt{R{T_{0}}}}}, \\
\hat{P}=\frac{P}{{{\rho _{0}}R{T_{0}}}},\text{ \ }({\hat{v}_{\alpha }},{\hat{%
u}_{\alpha }})=\frac{{({\nu _{\alpha }},}\text{ }{{u_{\alpha }})}}{\sqrt{R{%
T_{0}}}}, \\
(\hat{f},\text{ }{\hat{f}^{S}},\text{ }{\hat{f}^{eq}})=\frac{{(f,}\text{ }{{%
f^{S}},}\text{ }{{f^{eq}})}}{{{\rho _{0}}{{(R{T_{0}})}^{-D/2}}}}, \\
({\hat{f}_{i}},\text{ }\hat{f}_{i}^{S},\text{ }\hat{f}_{i}^{eq})=\frac{{({%
f_{i}},}\text{ }{f_{i}^{S},}\text{ }{f_{i}^{eq})}}{{{\rho _{0}}}}, \\
\hat{\kappa}=\frac{\kappa }{{{\rho _{0}}R{L_{0}}\sqrt{R{T_{0}}}}},\text{ \ }%
\hat{\mu}=\frac{\mu }{{{\rho _{0}}{L_{0}}\sqrt{R{T_{0}}}}}.%
\end{array}%
\right.
\end{equation}
Variables with ``$\wedge$'' on the left are dimensionless, while those on
the right are dimensional. For simplicity, the dimensionless symbol ``$%
\wedge $" is omitted in the rest of the paper.

\subsection{Viscous shock tube problem}

To evaluate the model's ability in capturing HNE
effects and its capacity in describing large-scale flow structures
with steep macroscopic gradients, we compare the second-order DBM simulation
results with Riemann solutions for various one-dimensional viscous shock
tube problems.

\subsubsection{Sod shock tube}

\begin{figure*}
\centering{%
\epsfig{file=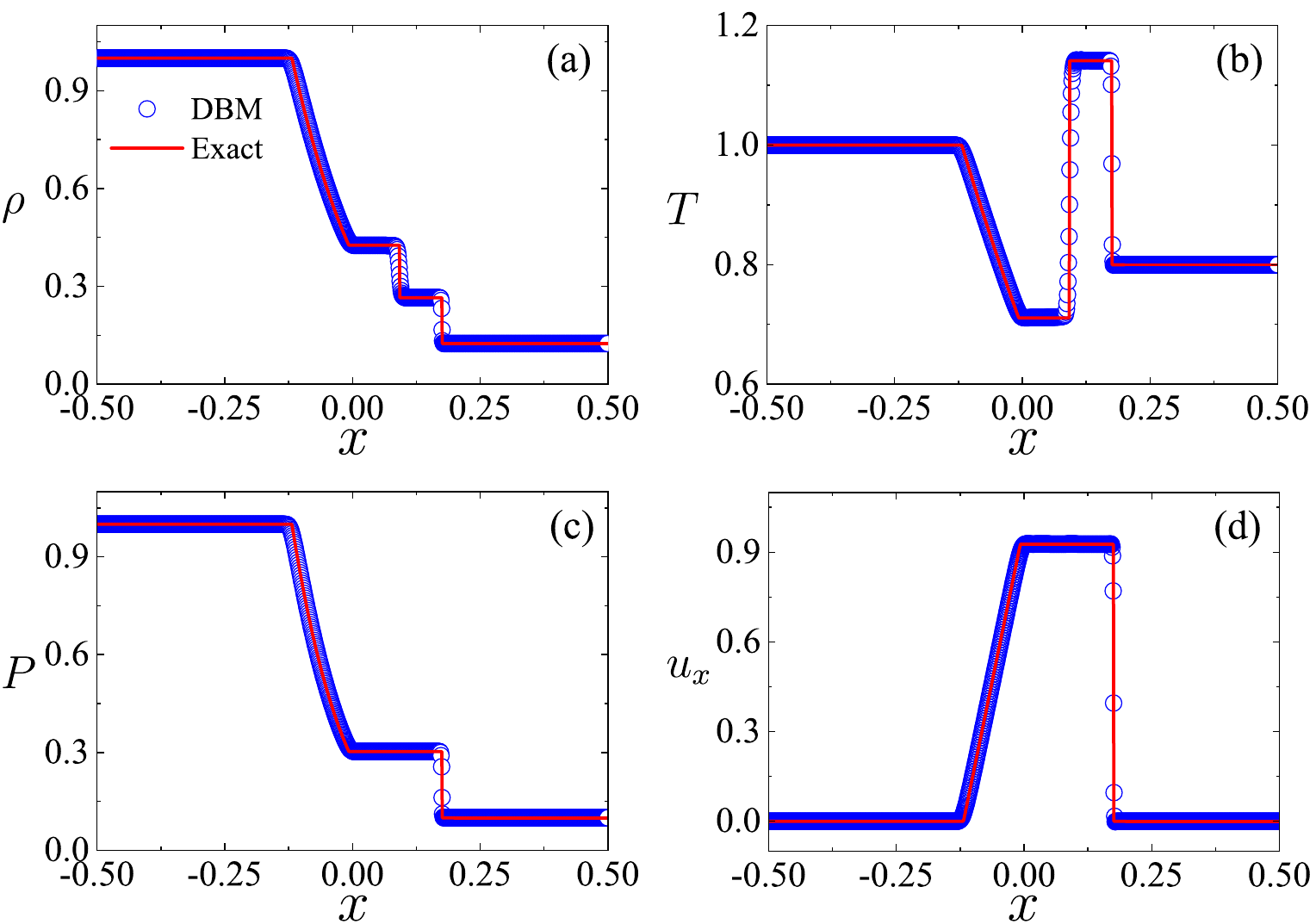,
			width=0.85\textwidth}}
\caption{Comparison of DBM simulation results with Riemann solutions for the
Sod shock tube problem at $t=0.1$.}
\label{sod}
\end{figure*}

The Sod shock tube problem~\citep{sod1978survey}, known for its rich characteristic structures,
is
commonly used to validate the effectiveness of fluid models. The initial
conditions for the left and right sides of the shock tube are as follows:
\begin{equation}
\left\{
\begin{array}{c}
{(\rho ,{u_{x}},{u_{y}},T){|_{L}}=(1.0,0.0,0.0,1.0)}, \\
{(\rho ,{u_{x}},{u_{y}},T){|_{R}}=(0.125,0.0,0.0,0.8)},%
\end{array}%
\right.
\end{equation}
where subscripts ``$L$" and ``$R$" denote the macroscopic variables on the left
and right sides of the discontinuity, respectively. The simulation parameters are $%
c = 1.0$, $\eta_0 = 1.0$; for $1 \leq i \leq 4$, $\eta_{ji} = i \eta_0 $;
for $13 \leq i \leq 16$, $\eta_{ji} = \eta_0 $; otherwise, $\eta_{ji} = 0$,
with $j=1, 2, 3$. The spatial steps are $\Delta x = \Delta y =
10^{-3}$, and the time step is $\Delta t = 2 \tau = 10^{-4}$. Additional
parameters include $\gamma = 1.4$ and $\Pr = 2.0$, with the grid size being $N_x
= N_y = 1000 \times 2$.
Fixed boundary conditions are imposed in the $x$%
-direction, and periodic boundary conditions are applied in the $y$%
-direction. Figure \ref{sod} compares DBM results with Riemann solutions for
density (a), temperature (b), pressure (c), and velocity (d) at $t = 0.1$.
The DBM effectively captures key structures, including rarefaction waves,
contact discontinuities, and shock waves, while significantly reduceding
numerical dissipation. The DBM results are in good agreement with the Riemann solutions.

\subsubsection{Mach 10 shock tube}

\begin{figure*}
\centering{%
\epsfig{file=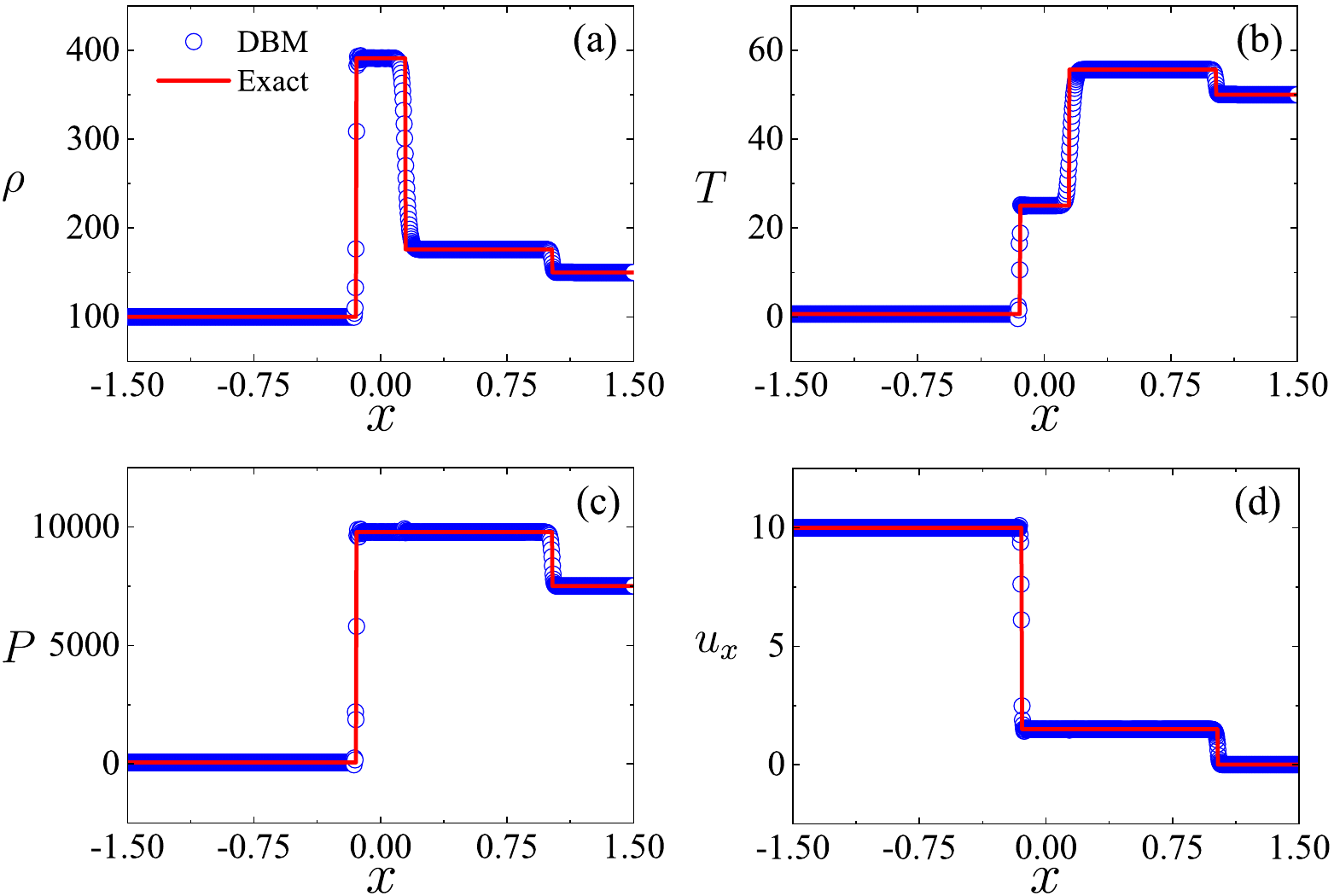,
			width=0.85\textwidth}}
\caption{Comparison of DBM simulation results with Riemann solutions for the
Mach 10 shock tube problem at $t=0.1$.}
\label{Mach10}
\end{figure*}

Next, a shock tube problem with a large temperature gradient and an initial
Mach number of $10$ is used to validate the model. The initial conditions are
\begin{equation}
\left\{ {%
\begin{array}{l}
{(\rho ,{u_{x}},{u_{y}},T){|_{L}}=(100.0,10.0,0.0,0.6)}, \\
{(\rho ,{u_{x}},{u_{y}},T){|_{R}}=(150.0,0.0,0.0,50.0)}.%
\end{array}%
}\right.   \label{eq22}
\end{equation}
The simulation parameters are as follows: $c=8.0$, $\eta _{0}=40.0$; for $%
1\leq i\leq 4$, $\eta _{1i}=i\eta _{0}$, and for $13\leq i\leq 16$, $\eta
_{1i}=\eta _{0}$; otherwise, $\eta _{i}=0$. $\Delta x=\Delta y=3\times
10^{-3}$, $\Delta t=2\tau =10^{-4}$, $\gamma =5/3$ and $\Pr =2.0$.
The simulation results at $t=0.1$ are shown in figure \ref{Mach10}. The DBM
numerical solutions closely match the Riemann solutions, clearly capturing both
the shock waves propagating both left and right, as well as the contact
discontinuity. Using the Rankine-Hugoniot relations, the propagation speeds
of the left and right shock waves are calculated as $u_{L}=-1.442$ and $%
u_{R}=10.189$, respectively.
The DBM simulation gives the following results:
$u_{L}=-1.439$ for the left shock wave and $u_{R}=10.157$ for the right
shock wave. The relative differences in propagation velocity for the
left and right shock waves are only $0.21\%$ and $0.31\%$, respectively. The
primary reason for these discrepancies is that the DBM accounts for TNE effects,
whereas the Riemann solution does not.
Viscous dissipation and heat flux reduce the shock wave propagation speed
and broaden the shock interface, which are not considered in the Riemann analytical
solution.

\subsubsection{Colella's explosion wave problem}

\begin{figure*}
\centering{%
\epsfig{file=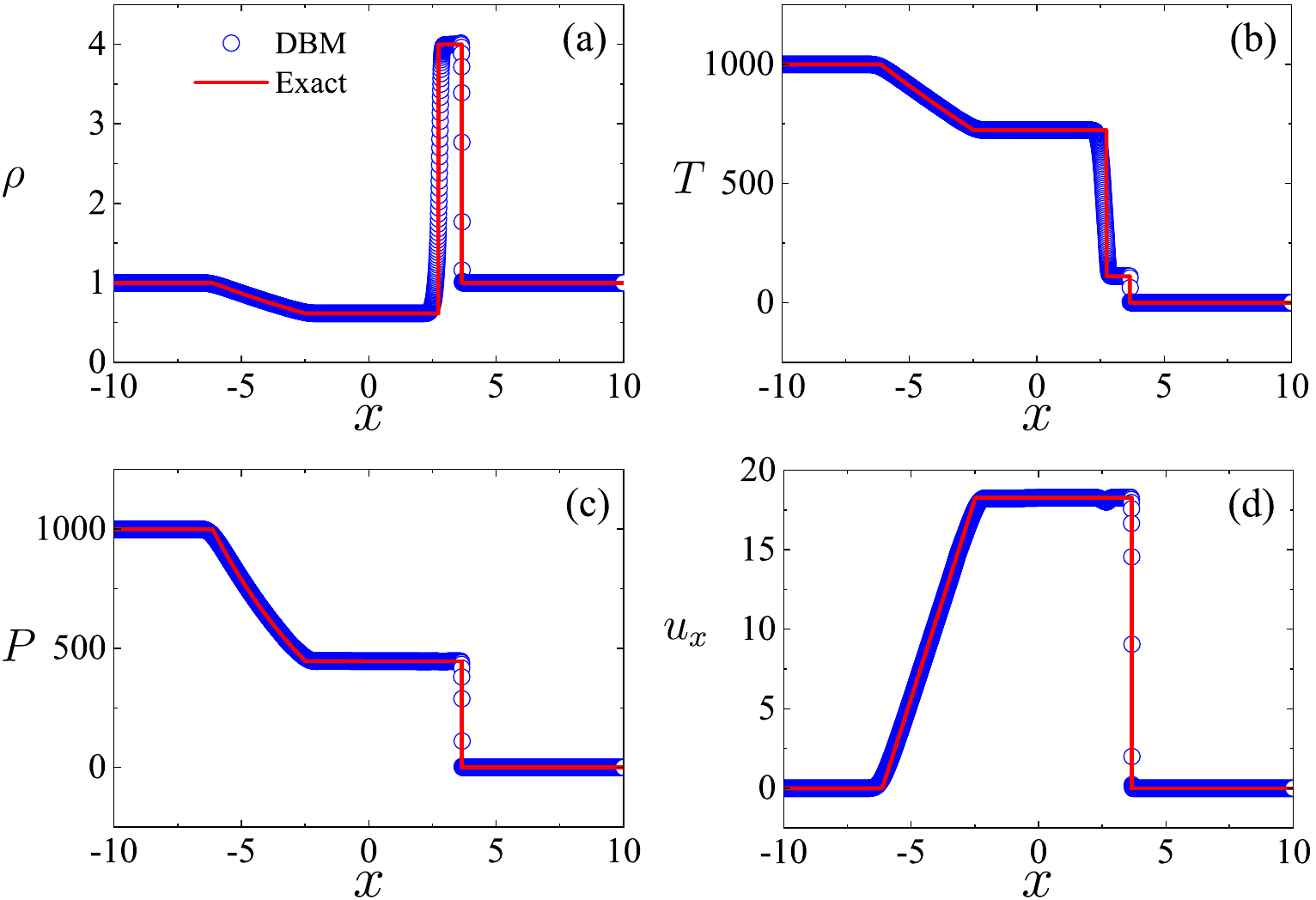,
			width=0.85\textwidth}}
\caption{Comparison of DBM simulation results with Riemann solutions for
Colella¡¯s explosion wave problem at $t=0.15$.}
\label{colella}
\end{figure*}

An explosion wave is a pressure wave generated by the rapid release of a large
amount of energy, with significant applications in defense engineering,
astrophysics, and civil safety. A typical
example of this phenomenon is Colella's explosion wave problem~\citep{woodward1984numerical},
which
serves as a classic test case characterized by extreme temperature and
pressure ratios, as well as high Mach numbers. The initial conditions for
the problem are
\begin{equation}
\left\{ {%
\begin{array}{c}
{(\rho ,{u_{x}},{u_{y}},T){|_{L}}=(1.0,0.0,0.0,1000.0)}, \\
{(\rho ,{u_{x}},{u_{y}},T){|_{R}}=(1.0,0.0,0.0,0.01)}.%
\end{array}%
}\right.  \label{eq23}
\end{equation}%
The model and simulation parameters are $c=18.0$, $\eta _{0}=40.0$; for $%
1\leq i\leq 4$, $\eta _{1i}=i\eta _{0}$, and for $13\leq i\leq 16$, $\eta
_{1i}=\eta _{0}$; otherwise, $\eta _{i}=0$. $\Delta x=\Delta y=5\times {%
10^{-3}}$, $\tau =2\Delta t={10^{-4}}$, $\gamma =5/3$, $\Pr =0.71$.
Figure \ref{colella} presents the simulation results at $t=0.15$. The DBM
numerical solution aligns well with the exact solution. The left-propagating
rarefaction wave, the contact discontinuity, and the strong
right-propagating shock wave are all accurately captured. The calculated
Mach number of the rightward shock wave is $Ma_{\text{DBM}}=189.562$,
compared to $Ma_{\text{RH}}=189.002$ from the Rankine-Hugoniot relations.
This negligible discrepancy is mainly due to the smaller relaxation time,
which directly affects the intensity of TNE.

\subsubsection{Astrophysical jet}

\begin{figure*}
\centering{%
\epsfig{file=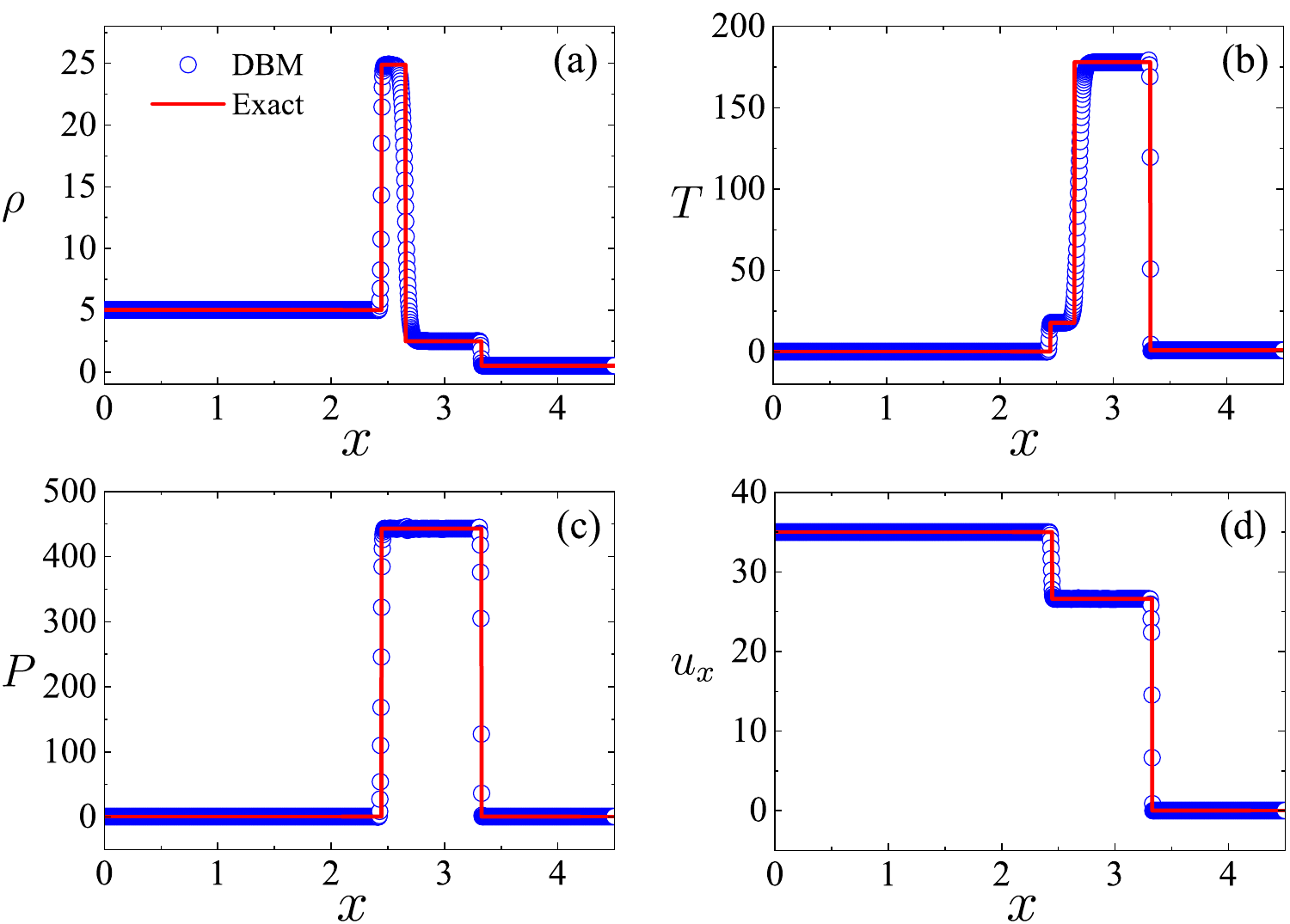,
			width=0.85\textwidth}}
\caption{Comparison of DBM simulation results with Riemann solutions for the
astrophysical jet problem at $t=0.1$.}
\label{jet}
\end{figure*}

Astrophysical jets are high-speed flows ejected from celestial bodies or
interstellar matter. These jets play a significant role in influencing the interstellar
medium, galaxy formation, and the dynamic evolution of the universe. The problem considered is a
shock tube problem with large
velocity gradients and multiple shock waves, characterized by high Mach
numbers~\citep{kallikounis2022particles,reyhanian2023exploring}. It has important applications in
areas auch as vortex engines,
rocket launches, rock explosions, and gas outbursts.
The initial conditions are
\begin{equation}
\left\{ {%
\begin{array}{c}
{\left( {\rho ,{u_{x}},{u_{y}},T}\right) {|_{L}}=(5.0,35.0,0.0,0.08254)} \\
{\left( {\rho ,{u_{x}},{u_{y}},T}\right) {|_{R}}=(0.5,0.0,0.0,0.8254)}%
\end{array}%
}\right.  \label{eq24}
\end{equation}%
The model and simulation parameters are $c=-14.0$, $\eta _{0}=20.0$; for $%
1\leq i\leq 4$, $\eta _{ji}=\eta _{0}$, and for $13\leq i\leq 16$, $\eta
_{ji}=(i-12)\eta _{0}$ with $j=1,2$; otherwise, $\eta _{i}=0$. $\Delta x=\Delta
y=3\times {10^{-3}}$, $\tau =3\times {10^{-5}}$, $\Delta t=5\times {10^{-4}}$
$\gamma =1.5$. 
The simulation results at $t=0.1$ are shown in figure \ref{jet}, demonstrating
that the DBM solution agrees well with the exact solution The two strong
shock waves moving rightward, along with the contact discontinuity are clearly
captured. The Rankine-Hugoniot relations yield $Ma_{1}=64.0892$ for the
first shock wave, and $Ma_{2}=30.286$ for the second shock wave. The DBM
solution gives $Ma_{1}=63.4475$ for the first shock wave, and $%
Ma_{2}=30.4059 $ for the second shock wave.

\subsubsection{Collision of two strong shock waves}

\begin{figure*}
\centering{%
\epsfig{file=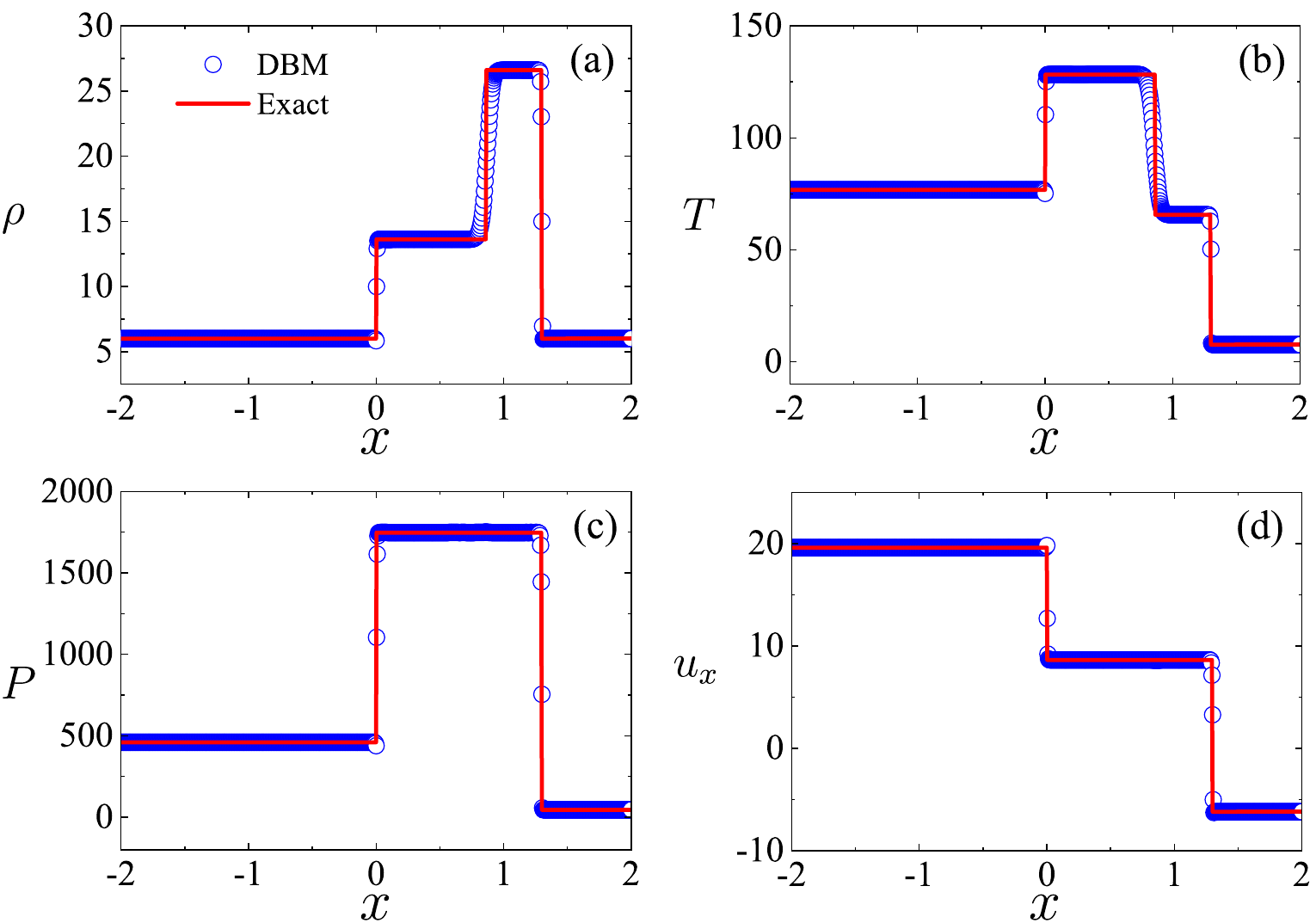,
			width=0.85\textwidth}}
\caption{Comparison of DBM simulation results with Riemann solutions for the
two strong shock collision problem at $t=0.1$.}
\label{collision}
\end{figure*}

The collision of strong shock waves is a critical problem in astrophysics,
plasma physics, explosion physics, and gas dynamics~\citep{gan2008two}. Studying these collisions
is crucial for
understanding natural phenomena and engineering processes, such as supernova
explosions, interactions between interstellar media, and inertial
confinement fusion experiments. Additionally, applications such as
high-energy particle experiments, vortex control, and jet engine design rely
on the a deep understanding of these interactions.
The initial conditions for this
problem are:
\begin{equation}
\left\{ {%
\begin{array}{c}
{\left( {\rho ,{u_{x}},{u_{y}},T}\right) {|_{L}}=(5.99924,19.5975,0,76.82540)%
} \\
{\left( {\rho ,{u_{x}},{u_{y}},T}\right) {|_{R}}=(5.99242,-6.19633,0,7.69222)%
}%
\end{array}%
}\right.  \label{eq25}
\end{equation}%
The model and simulation parameters are $c=9.0$, $\eta _{0}=20.0$. When $%
1\leq i\leq 4$, $\eta _{1i}=\eta _{2i}=\eta _{0}$, and when $13\leq i\leq 16$%
, $\eta _{1i}=\eta _{2i}=i\eta _{0}$; otherwise, $\eta _{i}=0$. $\Delta
x=\Delta y=5\times {10^{-3}}$, $\Delta =10\tau ={10^{-4}}$, $\gamma =1.5$, $%
\Pr =0.5$. 
The simulation results at $t=0.1$ are presented in figure \ref{collision}. The
two strong shock waves propagating to the left and right, as well as the
contact discontinuity, are accurately captured. The shock interfaces are
sharp, with no numerical oscillations. This demonstrates that DBM
effectively handles the complex dynamics of strong shock wave collisions.

\subsection{Simulation and analysis of thermodynamic nonequilibrium effects
at different orders}

The nonequilibrium behaviors in compressible flow systems are inherently complex and
multifaceted. Traditional fluid mechanics employs parameters such as the Knudsen number (Kn),
Mach number (Ma), Reynolds number (Re), viscosity, thermal conductivity, and macroscopic
gradients to characterize deviations from equilibrium. These parameters provide a coarse-grained,
condensed, and averaged representation of nonequilibrium effects from specific perspectives.
However, detailed information about nonequilibrium states, such as internal
energy distribution across different degrees of freedom, viscous stresses,
heat fluxes, fluxes of these quantities, higher-order non-conserved moments,
and their independent components, remains inaccessible and cannot be
directly analyzed using traditional metrics.
To address this limitation, a more comprehensive, multi-perspective, and finer-grained approach
is required to study nonequilibrium effects in compressible flow systems, complementing the
conventional descriptions provided by traditional fluid mechanics.
The DBM utilizes the nonequilibrium intensity vector, $\mathbf{S}_{\text{TNE}}$, to
realize a multi-perspective description of the nonequilibrium states and
effects. Among the components of $\mathbf{S}_{\text{TNE}}$,
viscous stress and heat flux, as key TNE measures in macroscopic fluid
dynamics equations, directly influence the direction and mode of system
evolution, necessitating detailed investigation.
The following section will validate DBM's ability to describe viscous stress and heat flux across
scales using specific examples.

\subsubsection{Typical thermodynamic nonequilibrium quantity: Viscous Stress}

Fluid collision problems involve interactions, intersections, or collisions
between two or more fluids. At collision points or regions, intense fluid
interactions give rise to pronounced local HNE and TNE effects.
 These effects manifest as abrupt macroscopic gradient shifts, sharp increases in viscous stress
 and/or heat flux, and the emergence of multiscale phenomena.
Investigating the multiscale behaviors and
nonequilibrium effects in fluid collisions is crucial for advancing our understanding of fluid
dynamics, mixing processes, and material exchange.

\begin{table*}
\caption{Physical parameters for cases with different orders of viscous
stress.}
\label{TableI}\centering
\begin{tabular}{ccccccc}
\toprule Case & $\Delta_{2xx}^*$ & Density & Temperature & Velocity &
Relaxation time & Pr \\
\midrule I & weak & $\rho_L=2\rho_R=2$ & $2T_L=T_R=2$ & $u_0 = 0.0$ & $%
3\times10^{-4}$ & 0.71 \\
II & moderate & $\rho_L=2\rho_R=2$ & $2T_L=T_R=2$ & $u_0 = 0.3$ & $%
3\times10^{-4}$ & 0.71 \\
III & strong & $\rho_L=2\rho_R=2$ & $2T_L=T_R=2$ & $u_0 = 1.0$ & $%
2\times10^{-3}$ & 0.71 \\
IV & super-strong & $10^{12}\rho_L=\rho_R=10^6$ & $2T_L=T_R=2$ & $u_0 = 0.3$
& $3\times10^{-3}$ & 0.71 \\
\bottomrule &  &  &  &  &  &
\end{tabular}%
\end{table*}

For the problem considered, the initial conditions are:
\begin{equation}
\rho (x,y) = \frac{{{\rho _L} + {\rho _R}}}{2} - \frac{{{\rho _L} - {\rho _R}%
}}{2}\tanh (\frac{{x - {N_x}\Delta x/2}}{{L_\rho }}),
\end{equation}
\begin{equation}
{u_x}(x,y) = - {u_0}\tanh (\frac{{x - {N_x}\Delta x/2}}{{L_u}}).
\end{equation}
For cases I-III in Table \ref{TableI}, the pressure is uniform $P_L = P_R=P_0$, $T(x,y)
= P_0/\rho(x,y)$. For case IV, the temperature is defined as $T(x,y) = \frac{%
{{T _L} + {T _R}}}{2} - \frac{{{T_L} - {T _R}}}{2}\tanh (\frac{{x - {N_x}%
\Delta x/2}}{{L_T }})$. For all cases, ${u_y}(x,y) = 0$. Here, $\rho_L$, $T_L$,
$P_L$ are the macroscopic quantities on the left side, while $\rho_R$, $T_R$,
$P_R$ are those on the right side of the fluids. $L_\rho$, $L_T$, and $L_u$
represent the interface widths for density, temperature and velocity,
respectively. $u_0$ is a free variable controlling the strength of the
collision. The other parameters are set as $\Delta x=\Delta y= 1.5 \times 10^{-3}$,
$\Delta t = 5 \times 10^{-5}$, $c=1.0$, $\eta_0=6.0$, and $\gamma=5/3$.

In an ideal gas system, the TNE behavior is governed by the
relaxation time $\tau$, as well as the macroscopic quantities and their
gradients. Consequently, we define a five-component vector, $\mathbf{S}_{\text{TNE}} = (\tau, %
\bm{\nabla} \rho, \bm{\nabla} T, \bm{\nabla} \mathbf{u}, \bm {\Delta}_{2}^*)$
to evaluate the system's nonequilibrium effects. By adjusting these
parameters, multiscale flows with varying nonequilibrium intensities and
Knudsen numbers can be generated.

\begin{figure*}
\centering{%
\epsfig{file=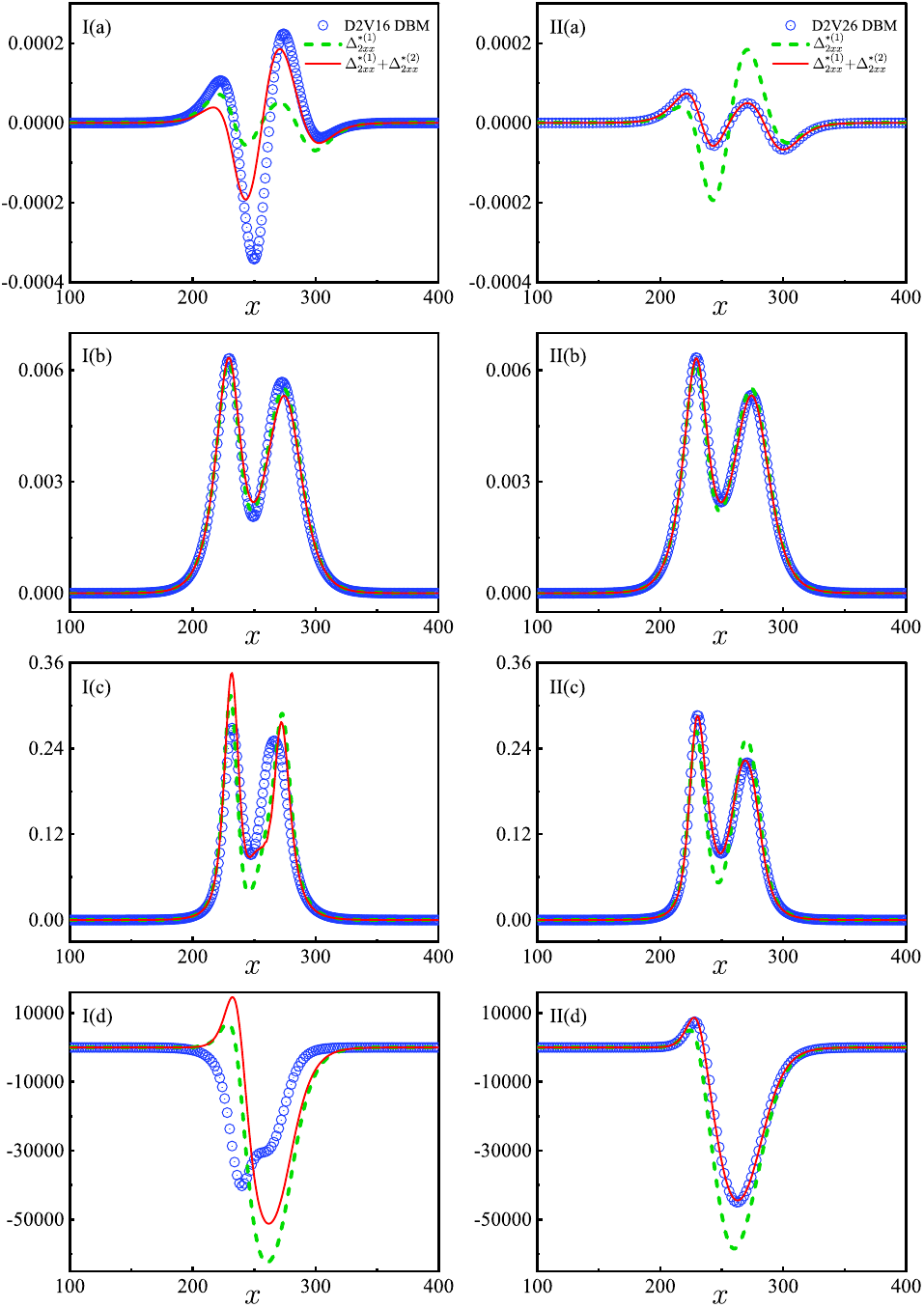,
			width=0.92\textwidth}}
\caption{Comparison of numerical solutions for viscous stress calculated
using the first-order DBM (left column) and the second-order DBM (right
column), for weak, moderate, strong, and super-strong TNE cases, where
dashed and solid lines represent the analytical solutions with first- and
second-order accuracies, respectively.}
\label{vis}
\end{figure*}

Figure \ref{vis} compares the numerical solutions of viscous stress obtained from the first-order
DBM (left column) and the second-order DBM (right column) under four different TNE intensities.
The results for the first three cases are taken at $t=0.025$, while the final case is taken at
$t=0.015$. In each panel, the analytical solutions with first- and second-order accuracies,
calculated from equations \ref{vis-1st} and \ref{vis-2nd}, are shown as dashed and solid lines,
respectively.
As shown in the first row of figure \ref{vis}, a smaller relaxation time $\tau$
and negligible velocity gradient $\bm{\nabla} \mathbf{u}$ lead to weaker TNE effects. At the
beginning of the evolution, where there are no velocity gradients ($u_0(x,y)=0$), the viscous
stress is driven by density and temperature gradients rather than velocity
gradients, as indicated by equation \ref{vis-2nd}, resulting in $\Delta_{2xx}^{*(2)} >
\Delta_{2xx}^{*(1)} \approx 0$. Subsequently, the
density and temperature gradient terms in equation \ref{vis-2nd} gradually
induce velocity gradients in equation \ref{vis-1st}, i.e., the
second-order TNE effects trigger the first-order TNE effects. As the system evolves,
$\Delta_{2xx}^{*(1)} $ surpasses $\Delta_{2xx}^{*(2)}$ and becomes the
dominant factor in the evolution.
Figure \ref{vis}(a) demonstrates that, even in weak nonequilibrium scenarios, the NS-level DBM,
which neglects second-order TNE effects, exhibits significant deviations between its analytical
and numerical solutions. In contrast, the Burnett-level DBM, which accounts for second-order TNE
effects, shows good agreement between analytical and numerical results, providing a more
comprehensive description of weak nonequilibrium effects.

Increasing the collision velocity $u_0$ strengthens the TNE effects, resulting in moderate
viscous stress (second row of figure \ref{vis}). At this stage, $\Delta_{2xx}^{(1)}$ is
approximately 30 times larger than in the case of weak TNE intensity. Importantly,
$\Delta_{2xx}^{(2)}$ can be neglected compared to $\Delta_{2xx}^{*(1)}$, indicating that velocity
gradients dominate the driving forces of TNE, rather than density and temperature gradients. Both
the second-order DBM (D2V26 model) and the first-order DBM (D2V16 model) effectively describe TNE
effects under this condition, although the second-order model provides better accuracy.

By further increasing $\tau$ and $u_0$, the TNE intensity is significantly
elevated to a strong level, as illustrated in the third row of figure \ref{vis}%
. Compared with the D2V16 model, the simulation results of the D2V26 model
consistently aligns with the second-order analytical solution, highlighting
the effectiveness and necessity of the second-order model under strong
nonequilibrium conditions. Moreover, in this scenario, at the left peak, $%
|\Delta_{2xx}^{*(1)}| \approx |\Delta_{2xx}^{*(1)} + \Delta_{2xx}^{*(2)}|$,
indicating that the second-order TNE exerts zero feedback on the first-order
nonequilibrium. In the center of the computational domain, $%
|\Delta_{2xx}^{*(1)}| > |\Delta_{2xx}^{*(1)} + \Delta_{2xx}^{*(2)}|$,
suggesting that second-order TNE provides negative feedback to
first-order TNE. At the right peak, $|\Delta_{2xx}^{*(1)}| <
|\Delta_{2xx}^{*(1)} + \Delta_{2xx}^{*(2)}|$, indicating that the
second-order TNE contributes positive feedback to the first-order TNE. It is
evident that higher-order TNE exhibits complex feedback effects on
lower-order TNE. The specific nature of the feedback depends on the local
flow states.

When the density ratio is further increased to $10^{12}$ and $\tau$ is
raised to $3 \times 10^{-3}$, the amplitude of $\Delta_{2xx}^*$ rapidly increases
to the order of $10^4$ [fourth row of figure \ref{vis}]. At this
point, the numerical solution of the D2V16 model deviates significantly from
both the first-order and second-order analytical solutions. The numerical
solution of the D2V26 model still perfectly matches the second-order analytical solution,
but significantly deviates from the
first-order one due to the contribution of the density gradient terms in
equation \ref{vis-2nd}.

Figure \ref{Kn}(a) illustrates the distributions of the Knudsen number $Kn$
calculated from gradients of macroscopic quantities in the case of
super-strong nonequilibrium. The Knudsen number is defined as $Kn=\lambda/L$%
, where $\lambda = c_s \tau$ represents the molecular mean-free-path and $L
= \phi / |\nabla \phi |$ is a local characteristic length scale, $c_s= \sqrt{%
\gamma T}$ is the local speed of sound, and $\phi$ signifies the macroscopic
quantity. Notably, the maximum Knudsen number $Kn_{\max}$ exceeds $0.3$,
which is far beyond the applicability of the NS model. In other words, the
D2V26 model is applicable for studying flow problems in the early transition
flow regime.

From figure \ref{vis}, we conclude that whether or not to consider the
higher-order TNE effects depends on their relative importance compared to
lower-order TNE effects, or the discrepancies between analytical solutions
of different orders, rather than solely on the Knudsen number. For example,
second-order TNE effects should be considered in rows 1, 3, and 4, even
though the maximum local Knudsen numbers for the weak and strong
nonequilibrium cases are only $5.18 \times 10^{-3}$ and $0.09$,
respectively. Therefore, the relative nonequilibrium intensity is used to
measure the relative importance of higher-order nonequilibrium effects,
defined as $R_{\text{TNE}} = |\bm{\Delta}_{m,n}^{*(j+n)}/\bm{\Delta}%
_{m,n}^{(j)}|$. In the scenarios shown in rows 1, 3, and 4, the maximum
relative nonequilibrium intensities $R_{\text{TNE-max}} =
\Delta_{2xx}^{*(2)}/\Delta_{2xx}^{*(1)}$ are $0.73$, $0.78$, and $0.24$,
respectively. This approach provides a more reasonable method for describing
the multiscale characteristics under investigation and for determining the
appropriate model order in physical modeling.

\subsubsection{Typical thermodynamic nonequilibrium quantity: Heat Flux}

\begin{table*}
\caption{Physical parameters for cases with different orders of heat flux.}
\label{TableII}\centering
\begin{tabular}{ccccccc}
\toprule Case & $\Delta_{3,1x}^*$ & Density & Temperature & Velocity &
Relaxation time & Pr \\
\midrule I-III & \makecell{weak \\ moderate \\ strong} & $2\rho_L=\rho_R=2$
& $T_L=T_R=1.5$ & \makecell{$u_{xL} = 0.2$ \\ $u_{xR} = 0$} & $%
2\times10^{-3} $ & 0.3 \\
IV & super-strong & $10^{12}\rho_L=\rho_R=10^6$ & $2T_L=T_R=2$ & %
\makecell{$u_{xL} = 0.5$ \\ $u_{xR} = 0$} & $2.5\times10^{-3}$ & 0.6 \\
\bottomrule &  &  &  &  &  &
\end{tabular}%
\end{table*}

\begin{figure*}
\centering{%
\epsfig{file=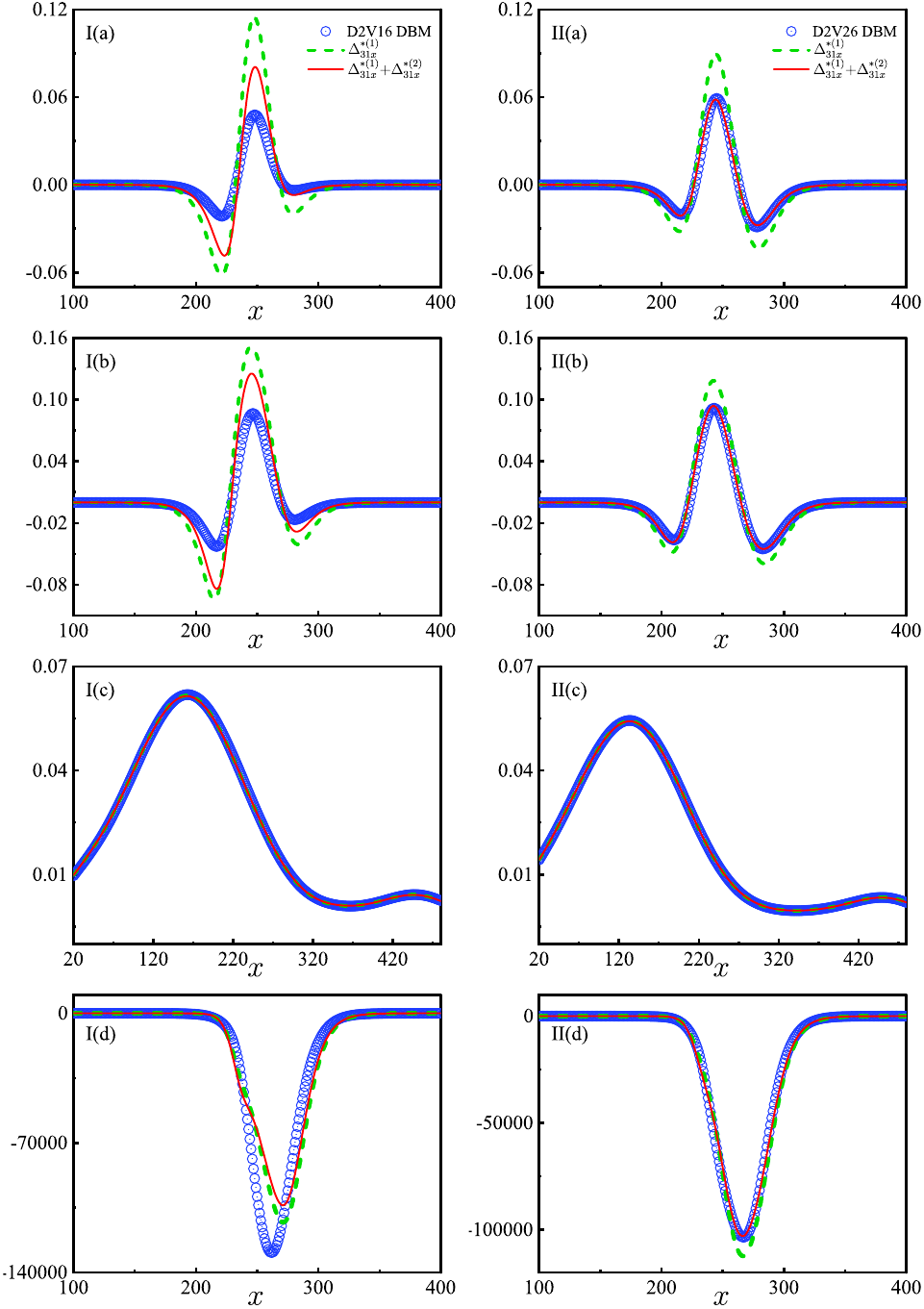,
			width=0.92\textwidth}}
\caption{Comparison of numerical solutions for heat flux calculated using
the first-order DBM (left column) and the second-order DBM (right column)
under weak, moderate, strong, and super-strong TNE cases, where dashed and
solid lines represent the analytical solutions with first- and second-order
accuracies, respectively.}
\label{heat}
\end{figure*}

Next, the model's ability to describe multiscale heat flux is examined similarly. The initial
conditions are given by
\begin{equation}
\phi (x,y) = \frac{{{\phi_L} + {\phi_R}}}{2} - \frac{{{\phi_L} - {\phi_R}}}{2%
}\tanh (\frac{{x - {N_x}\Delta x/2)}}{{L_\phi }}),  \label{e30}
\end{equation}
where $\phi= [\rho, T, u_x]$. Table \ref{TableII} lists the initial
macroscopic parameters on the left and right sides of the system for four
cases. Figure \ref{heat} compares the numerical solutions from the
first-order DBM (left column) and second-order DBM (right column) across
four levels of nonequilibrium intensity, along with corresponding
analytical solutions for both orders.

The first three rows of figure \ref{heat} depict weak, strong, and moderate
nonequilibrium cases at $t=0.0175$, $0.025$, and $0.5$, representing distinct stages of the
evolution process. Initially, in the absence of a
temperature gradient ($T_L = T_R = 1.5$), the heat flux is primarily driven
by density and velocity gradients, as described in equation \eqref{heat-2nd}.
This results in $\Delta_{3,1x}^{(2)} > \Delta_{3,1x}^{(1)} \approx 0$.
Subsequently, the density and velocity gradients gradually induce
temperature gradients, leading to $\Delta_{3,1x}^{*(1)}$ surpassing $%
\Delta_{3,1x}^{*(2)}$, as shown in the second and third rows of figure \ref%
{heat}.  Specifically, in panel IIC, $\Delta_{3,1x}^{*(2)}$ becomes negligible
relative to $\Delta_{3,1x}^{*(1)}$ as the density and velocity gradients
diminish.  These findings demonstrate that in an evolving nonequilibrium
system, the nonequilibrium intensity varies substantially across different stages.
Initially, one or more nonequilibrium driving forces initiate
additional types of nonequilibrium effects, which at a certain point cause
the nonequilibrium effects to peak, forming distinct multiscale
structures within the system. As the system evolves, macroscopic quantities
become more uniform, and the nonequilibrium driving forces weaken due to
dissipation, leading to a simplified and stable flow structure.
As a multiscale model, it should adapt to to capture cross-scale nonequilibrium
effects. The results of the D2V26 model, displayed in the right column of figure \ref%
{heat}, precisely exemplify this capability.

When further increasing the density, temperature, and velocity gradients,
along with relaxation time, the heat flux transitions into the super-strong
regime, as depicted in the fourth row of figure \ref{heat}. At this stage, the
low-order model fails to adequately capture such intense nonequilibrium effects, whereas
the second-order model's numerical solution still matches the second-order
analytical solution.

\begin{figure*}
\centering{%
\epsfig{file=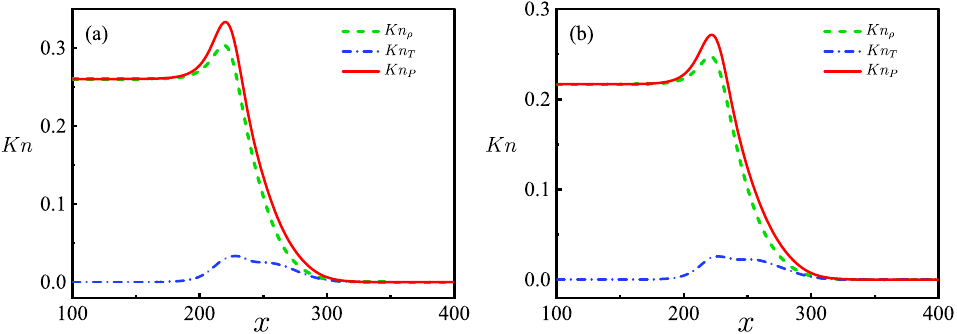,
			width=0.96\textwidth}}
\caption{Distributions of the Knudsen number calculated from gradients of macroscopic
quantities in the super-strong viscous stress (a) and heat flux (b) cases.}
\label{Kn}
\end{figure*}

Additionally, as illustrated in figure \ref{Kn}(b), the
maximum Knudsen number, derived from gradients of macroscopic quantities in the
super-strong nonequilibrium case, exceeds $0.25$.
interactions and intensified nonequilibrium forces that emerge in the super-strong regime,
providing a more physical description of the system's evolving heat flux.
This observation underscores the superior capability of the second-order model to address the
complex multiscale interactions and intensified nonequilibrium forces characteristic of the
super-strong regime, thereby providing a more accurate and comprehensive description of the
system's evolving heat flux.

From the above analysis of a multiscale nonequilibrium flow problem, the following conclusions
can be drawn:
(1) The first-order model
overestimates nonequilibrium effects under both weak and strong
nonequilibrium conditions;
(2) The D2V26 model, which satisfies more kinetic moment relations and provides a comprehensive
description of first- and second-order TNE, accurately predicts nonequilibrium effects consistent
with analytical solutions in the three analyzed scenarios. This highlights its multiscale
capability in describing nonequilibrium phenomena.
(3) In nonequilibrium flows, the consideration of which-order
nonequilibrium effects to include in physical modeling depends on the relative
nonequilibrium intensity rather than the Knudsen number;
(4) In most cases shown in the figure, second-order nonequilibrium effects oppose first-order
effects, resulting in a negative feedback.

\section{Thermodynamic nonequilibrium effects and entropy production
mechanisms in the regular reflection of shock wave}

\begin{figure}
\centering{%
\epsfig{file=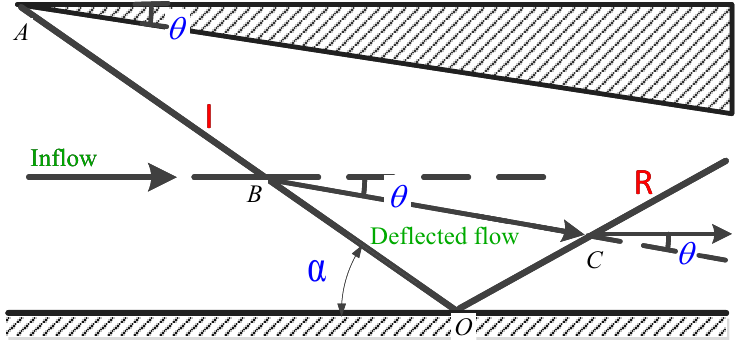,
			width=0.55\textwidth}}
\caption{Schematic diagram of the regular shock wave reflection.}
\label{schem}
\end{figure}

\begin{figure}
\centering{%
\epsfig{file=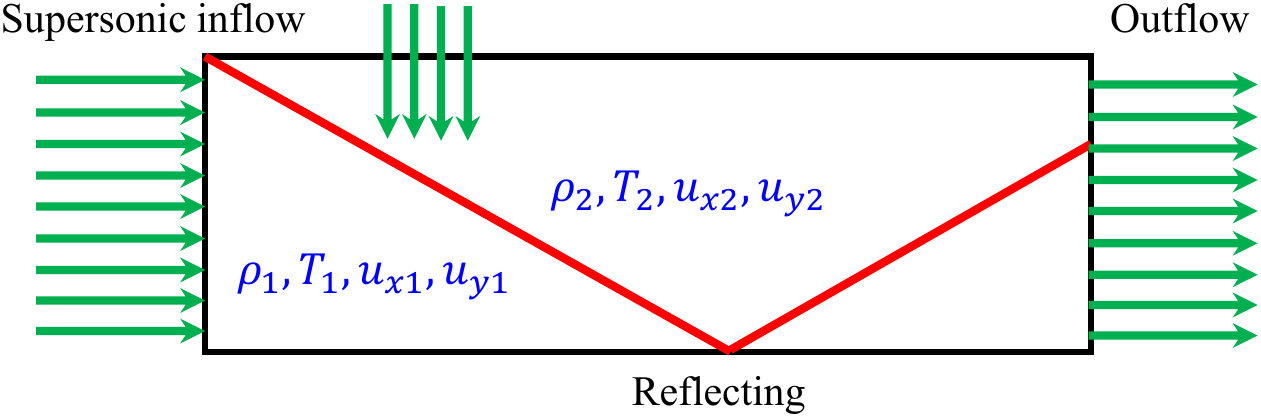,
			width=0.55\textwidth}}
\caption{Boundary condition settings for regular shock wave reflection.}
\label{BC}
\end{figure}

This section investigates the TNE effects and entropy production mechanisms in
regular shock wave reflection, which are crucial for understanding the
complex flow dynamics and wave structures in a supersonic inlet.

The supersonic inlet of a scramjet engine, a vital aerodynamic component,
exhibits flow phenomena including oblique shocks induced by the inlet
compression angle, boundary layers generated by flow-wall interactions,
separation bubbles caused by reflected shock-boundary layer interactions,
reflected shocks near the inlet lip, and expansion waves at
corners~\citep{li1997parametric,huang2020recent,chang2017recent,qiu2017lattice,bao2022study}.
These phenomena collectively increase aerodynamic drag, reduce air mass flow rate, and decrease
inlet efficiency, ultimately impacting the engine's stability, reliability, and overall
performance.

TNE effects and entropy production mechanisms play a vital role in governing
interactions among shocks, boundary layers, and separation bubbles.
Shock wave reflection, in particular, significantly influences wave drag and
frictional drag, which are the main contributors to the inlet's total aerodynamic drag.
Analyzing the thermodynamic properties and flow dynamics of
shock wave reflection reveals mechanisms underlying performance
losses and aids in developing strategies to minimize drag, ultimately enhancing thrust and
combustion efficiency in supersonic
engines.

Li \emph{et al.} investigate the transition from regular reflection to Mach reflection  in steady supersonic flow, caused by an upstream disturbance~\citep{li2011time}.
Through computational fluid dynamics simulation, they observe multiple shock interactions in the early stage, including a triple-shock structure and shock/slipline interaction. 
Yao \emph{et al.} address shock reflection in supersonic and hypersonic intake flow under off-design conditions ~\citep{yao2013shock}. They focus on the interaction between the incident shock wave, an upstream expansion wave generated by the lip, and the downstream shock wave interacting with a cowl-turning deflected shock wave. The study also analyzes the influence of the lip and cowl turning angles on the flow structure and the transition between regular and Mach reflection. The findings suggest that the interference between the expansion and shock waves significantly alters the dual-solution domain, with the Mach stem height increasing with the turning angle.

As research progresses, an increasing number of studies focus on the nonequilibrium effects in shock reflection, such as viscosity, heat conduction, and rarefaction effects.
Khotyanovsky  \emph{et al.} investigate the viscous effects in steady shock wave reflection for a Mach 4 flow of a monatomic gas, using both NS and DSMC simulations~\citep{khotyanovsky2009viscous}. They focus on the impact of viscosity and heat conduction near the shock intersection, where flow parameters deviate from inviscid predictions. The study highlights the discrepancies caused by viscous and heat transfer effects, which are significant in high-speed aerodynamics, particularly during shock interactions in supersonic and hypersonic flows.
Shoev \emph{et al.} explore the effects of viscosity and rarefaction on regular shock wave reflections~\citep{shoev2017stationary}. They highlight that traditional Rankine-Hugoniot relations describe flow parameters behind shock waves but do not provide information on the internal structure of the shock wave itself.
For regular shock reflection in a viscous gas, they present an analytical solution in the vicinity of the reflection point. This solution consists of three zones with constant parameters, separated by infinitely thin discontinuity lines (incident and reflected shock waves). In the case of a viscous gas, the shock waves have finite thickness, and the viscous effects must be considered.
Timokhin \emph{et al.} extend the classical Mott-Smith solution for one-dimensional normal shock wave structure to the two-dimensional regular shock reflection problem~\citep{timokhin2022mott}. The solution is expressed in terms of a nonequilibrium molecular velocity distribution function along the symmetry-plane streamline, represented as a weighted sum of four Maxwellians.
The applicability of the solution is analyzed using DSMC simulations across a range of incident shock wave intensities. The results show that the accuracy of the solution improves with increasing Mach number, and becomes especially accurate for strong shocks where $Ma > 8$.
Qiu and Bao \emph{et al.} study the nonequilibrium characteristics during the development of regular reflections of shock wave \citep{qiu2017lattice,Qiu-POF-2020,bao2022study,qiu2024mesoscopic}.
 They point out that the nonequilibrium kinetic moments related to mass can be used to determine the position of boundary layer separation, while nonequilibrium kinetic moments related energy determine the total energy variation inside the boundary layer \citep{bao2022study}.
However, the nonequilibrium effects in the regular reflection process are rich and complex. Viscosity and heat conduction alone are insufficient to capture their intricate characteristics. Therefore, in this section, we use the nonequilibrium intensity vector, $\mathbf{S}_{\text{TNE}}$, defined by DBM to meticulously describe the discrete/non-equilibrium states, effects, and behaviors during the shock reflection process.

Figure \ref{schem} shows a schematic diagram of the two-dimensional regular shock wave reflection
flow field.
An inclined wedge with an angle of $\theta$ is positioned at the upper section of the
flow field.
The incoming flow with a high Ma propagates from left to right, forming
an oblique shock with an incident angle of $\alpha $ under the influence of
the wedge.
When the incident angle is small enough, the shock reflects off the lower wall and  a reflected shock is formed, dividing the region between the wedge and the lower wall into three parts.
The two sides of the incident shock wave (region 1 and 2) satisfy the following relationship
\begin{equation}
\left\{
\begin{array}{l}
\frac{{{\rho _{2}}}}{{{\rho _{1}}}}=\frac{{\left( {\gamma +1}\right)
M_{1}^{2}{{\sin }^{2}}\alpha }}{{2+\left( {\gamma -1}\right) M_{1}^{2}{{\sin
}^{2}}\alpha }} \\
\frac{{{p_{2}}}}{{{p_{1}}}}=\frac{{2\gamma M_{1}^{2}{{\sin }^{2}}\alpha
-\left( {\gamma -1}\right) }}{{\gamma +1}} \\
M_{2}^{2}{\sin ^{2}}\left( {\alpha -\theta }\right) =\frac{{\gamma +1+\left(
{\gamma -1}\right) (M_{1}^{2}{{\sin }^{2}}\alpha -1})}{{\gamma +1+2\gamma
(M_{1}^{2}{{\sin }^{2}}\alpha -1)}} \\
\tan \theta =\tan \alpha \frac{{M_{1}^{2}{{\cos }^{2}}\alpha -{{\cot }^{2}}%
\alpha }}{{1+\frac{1}{2}M_{1}^{2}\left( {\gamma +\cos 2\alpha }\right) }}%
\end{array}%
\right. .
\end{equation}%
Figure \ref{BC} depicts the boundary conditions applied in the simulation.
The left boundary employs a supersonic inflow boundary condition, while the
right boundary adopts an outflow boundary condition.
The upper and lower boundaries adopt the Dirac and the specular reflection conditions, respectively.
The Shakhov-BGK DBMs considering the 1st and 2nd orders of TNE are used for the simulation, respectively.
The initial inflow is uniformly set as $(\rho
,T,u_{x},u_{y})=(\rho _{0},T_{0},Ma\sqrt{\gamma T_{0}},0.0)$, with model
parameters $\tau =10^{-3}$, $\Delta x=\Delta y=0.005$, $c=1.6$. For $1\leq
i\leq 4$, $\eta _{i}=\eta _{0}i$, and for $13\leq i\leq 20$, $\eta _{i}=1.6$%
, while in other cases $\eta _{i}=0.0$.

\subsection{Necessity of high-order models: Ensuring positivity of entropy
production}

The origin of entropy production primarily stems from irreversible processes within the system.
Using the DBM, entropy production rate due to both viscosity and heat
flux can be obtained simultaneously.
The total entropy production rate and the contributions from viscosity and heat flux are defined
as follows:
\begin{equation}
\dot S = \frac{{dS}}{{dt}} = \int ( {\bm{\Delta }}_{3,1}^* \cdot \nabla \frac{%
1}{T} - \frac{1}{T}{\bm{\Delta }}_2^*:\nabla \mathbf{u})d\mathbf{r}
\end{equation}
\begin{equation}
{\dot S_{\mathrm{NOMF}}} = \frac{{d{S_{\mathrm{NOMF}}}}}{{dt}} = - \int {%
\frac{1}{T}{\bm{\Delta }}_2^*:\nabla \mathbf{u}} d\mathbf{r}
\end{equation}
\begin{equation}
{\dot S_{\mathrm{NOEF}}} = \frac{{d{S_{\mathrm{NOEF}}}}}{{dt}} = \int {{%
\bm{\Delta }}_{3,1}^*} \cdot \nabla \frac{1}{T}d\mathbf{r}.
\end{equation}
When performing physical modeling, the accuracy of the required DBM depends
on the specific nature of the problem and research requirements.
In cases
with weak nonequilibrium effects, a first-order DBM is sufficient to meet
the physical requirements. However, with increased system discretization and nonequilibrium
effects, higher-order DBMs that account for
higher-order TNE effects can provide more reasonable results.
In this study, the required order of DBM for studying shock-induced regular reflection
is determined by examining the positivity of entropy production.
Specifically, the initial conditions are ste as $(\rho, T, u_x, u_y) = (1.0,
1.0, 2.2\sqrt{1.4}, 0.0)$.

Figure \ref{CompareEntropy} shows entropy production rates using
 first-order (a) and second-order (b) DBMs along $y = 40\Delta y$
at $t=1.8$, with lines in (b) showing second-order analytical
solutions.
It can be observed that, in (a), the curves for the three types
of entropy production rates calculated from the first-order DBM exhibit
significant numerical oscillations, including substantial negative values near
the incident and reflected shocks.
In contrast, in (b), the entropy production rate curves calculated from the second-order
DBM are smooth, consistently positive, and closely match the
second-order analytical solutions.
This indicates that the second-order model, compared to the first-order model, more accurately
represents viscous stress and heat flux, captures dissipative mechanisms precisely, ensures
entropy production non-negativity in line with the second law of thermodynamics, and exhibits
better numerical stability, making it more suitable for analyzing strong nonequilibrium flows.

\emph{This demonstrates that, compared to the first-order model, the second-order model provides
more accurate representations of viscous stress and heat flux, describes dissipative mechanisms
more precisely, ensures the non-negativity of entropy production according to the second law of
thermodynamics, and exhibits better numerical stability, making it more suitable for studying
strong nonequilibrium flows}.

\subsection{Process of the regular reflection of shock wave}

\begin{figure*}
\centering{%
\epsfig{file=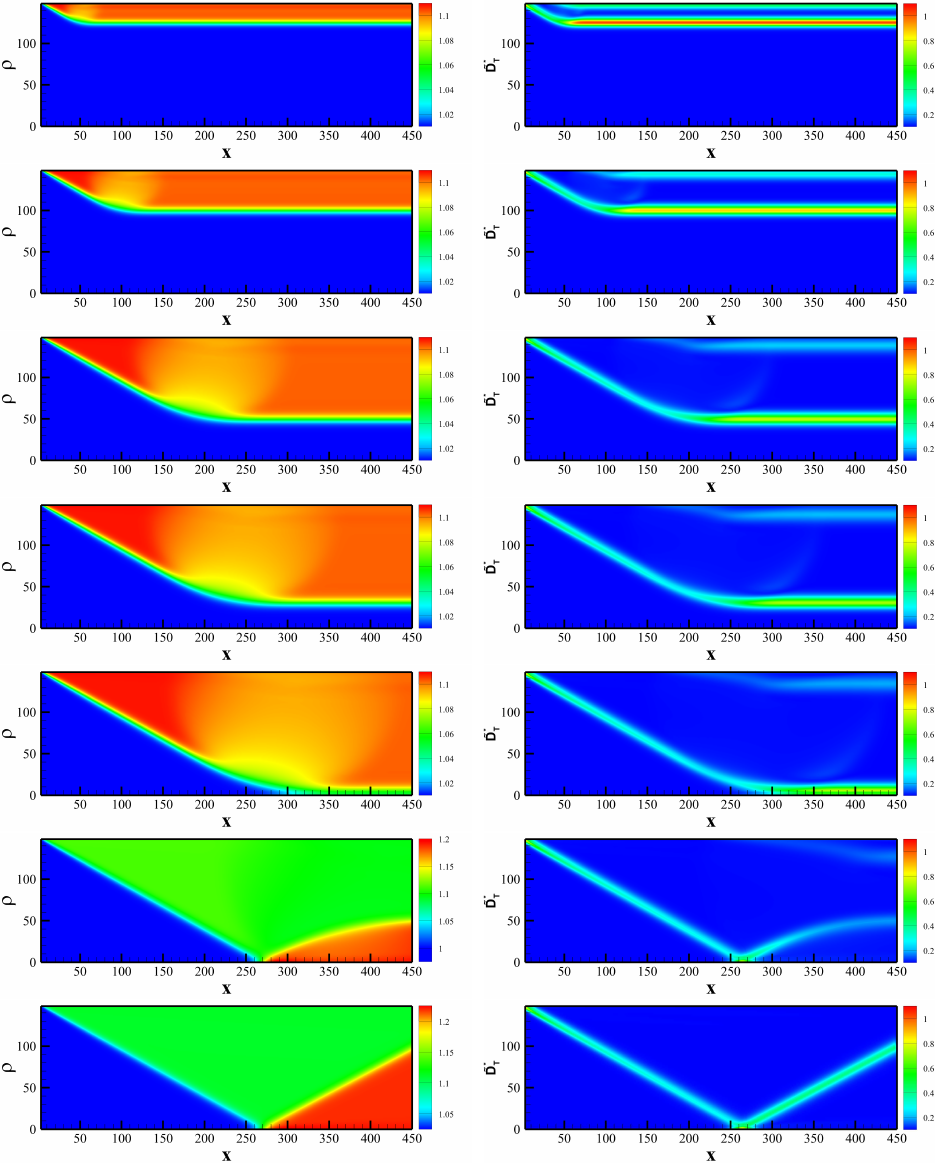,
			width=0.92\textwidth, clip}}
\caption{ Density contours (left column) and nonequilibrium intensity
contours (right column) at characteristic moments during the regular shock wave
reflection process for $Ma=2.2$. The corresponding moments, from top to
bottom, are $t=0.1, 0.2, 0.4, 0.475, 0.575, 0.8, 1.8$, respectively.
}
\label{RR}
\end{figure*}
Depending on the incident angle, the reflection of an oblique shock on a
plane can be categorized as regular reflection or Mach reflection.
Figure \ref{RR} shows the density contours (left column) and nonequilibrium intensity contours (right column) at characteristic moments during the regular reflection process of a shock wave for $Ma=2.2$.
It can be seen that the regular reflection process can be divided into three stages:
(1) When $t<0.575$, the left-incoming flow is compressed by the upstream flow, forming a transient curved incident shock with positive curvature that gradually propagates downward.
(2) At about $t=0.575$, the incident shock strikes the lower wall and reflects, creating a curved reflected shock with negative curvature~\citep{shi2020second}. During $0.575<t<1.8$,
the curved reflected shock gradually straightens under the influence of the lower wall.
(3) When $t>1.8$, the reflected shock stabilizes, both the incident and reflected shocks become clearly distinguishable.

By measuring the incident or reflected shock angles, the model's ability to
capture strong nonequilibrium shocks can be validated.
When the regular reflection develops to a steady state, the incident angle of the shock simulated by BDM is $29.1^\circ$ as shown in figure \ref{RR}, with a relative error of $0.34\%$ compared to the analytical value.
This confirms the accuracy of the DBM in capturing the shock incident angle. As
shown in the right column of figure \ref{RR}, nonequilibrium effects are
concentrated near the shock interfaces, while the rest of the flow field
remains close to equilibrium. Additionally, the intensity of the reflected shock
is slightly stronger than that of the incident shock, with an intensity ratio
of approximately $1.41$ at $t = 1.8$. This difference arises from the relatively lower Mach
number of the upstream flow. By analyzing the nonequilibrium intensity, the positions and
evolution characteristics of the incident and reflected shock waves can be effectively
determined.

\begin{figure*}
\centering{%
\epsfig{file=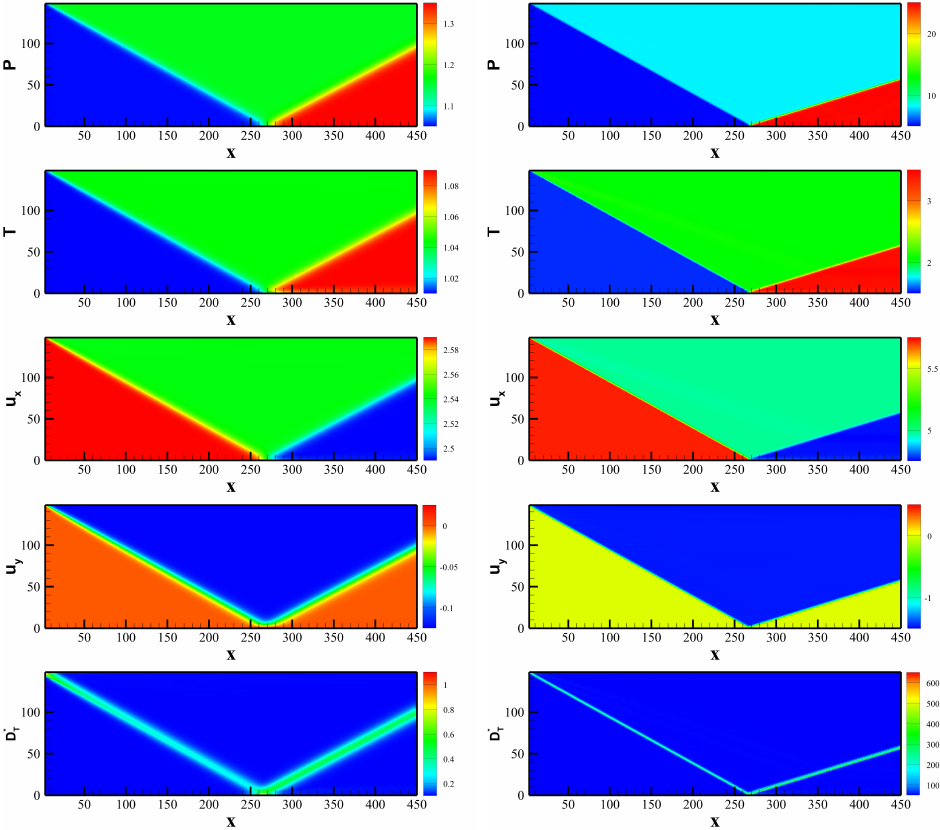,
			width=0.92\textwidth}}
\caption{Density, temperature, pressure, $x$-velocity, $y$-velocity, and
nonequilibrium intensity distributions at steady state for the regular
shock wave reflection (left column: $Ma = 2.2$, right column: $Ma = 5.0$). }
\label{Ma2-5}
\end{figure*}

Figure \ref{Ma2-5} presents the steady-state distributions of density,
temperature, pressure, $x$-velocity, $y$-velocity, and nonequilibrium
intensity for the regular shock wave reflection at $Ma = 2.2$
(left column) and $Ma = 5.0$ (right column).
When $t > 1.8$, the shock reflection process reaches a steady state. In both cases, the incident
and
reflected shocks divide the flow field into three regions, displaying no
oscillations but maintaining a finite interface width. This suggests that the
second-order model effectively captures large-scale flow structures in supersonic
flows. Compared to the $Ma = 2.2$ case, the shock interface in the $Ma=5.0$ case is significantly
narrower, indicating an enhanced compressive effect on the airflow at higher Mach numbers.
Furthermore, the total nonequilibrium
strength in the $Ma = 5.0$ case is much higher than in the $Ma = 2.2$ case,
with a ratio of approximately $6 \times 10^2$. This indicates that higher
Mach numbers can rapidly increase the system's nonequilibrium strength.

\subsection{Evolution of thermodynamic nonequilibrium quantities}

\begin{figure*}
\centering
\begin{subfigure}[b]{0.92\textwidth}
        \epsfig{file=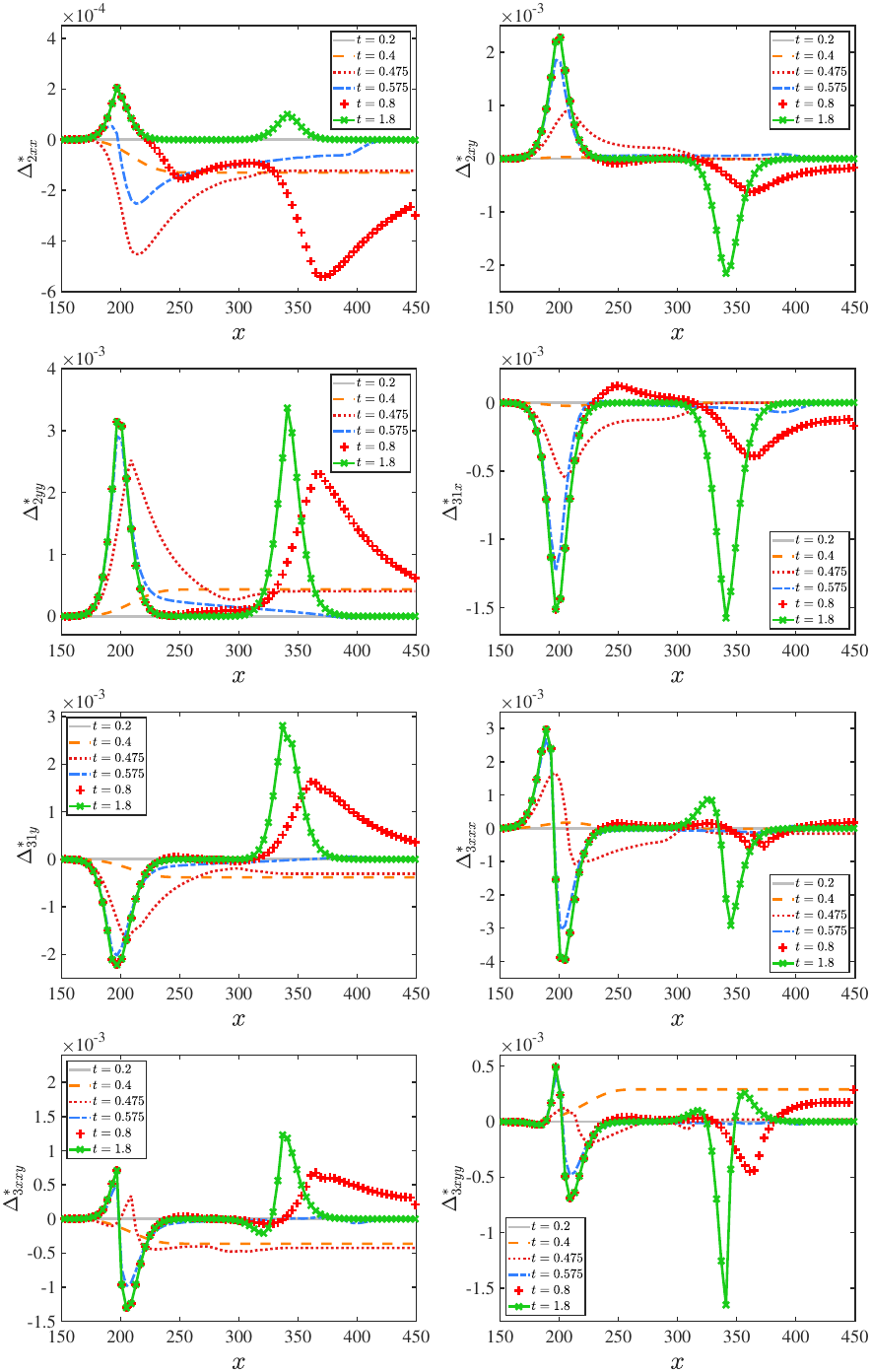, width=\textwidth}
        \caption{Continue.}
        \label{fig:sub1}
    \end{subfigure}
\label{RRT1}
\end{figure*}
\begin{figure*}
\centering
\begin{subfigure}[b]{0.92\textwidth}
        \epsfig{file=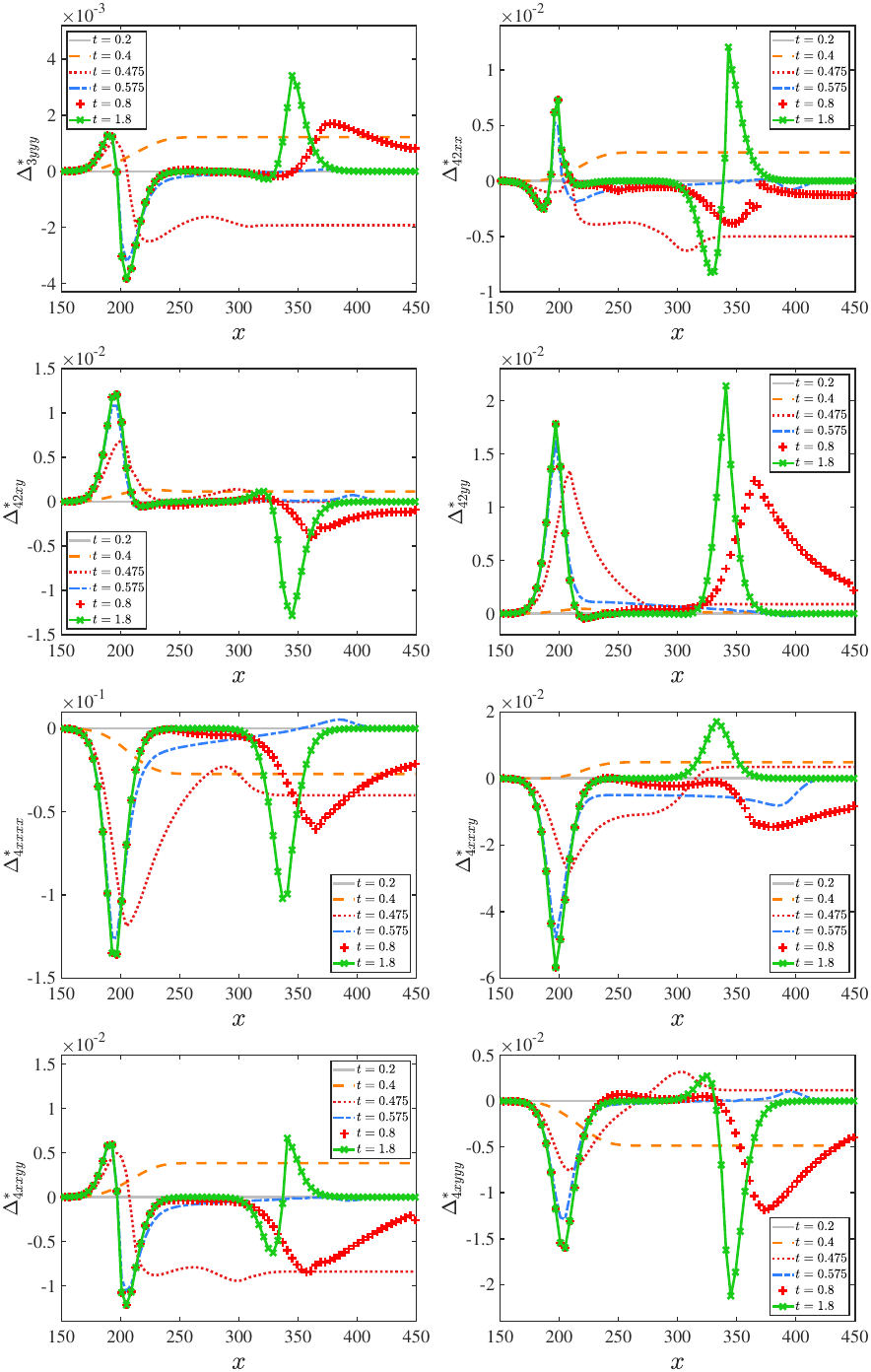, width=\textwidth}
        \caption{Continue.}
        \label{fig:sub2}
    \end{subfigure}
\label{RRT2}
\end{figure*}
\begin{figure*}
\centering
\begin{subfigure}[b]{0.92\textwidth}
        \epsfig{file=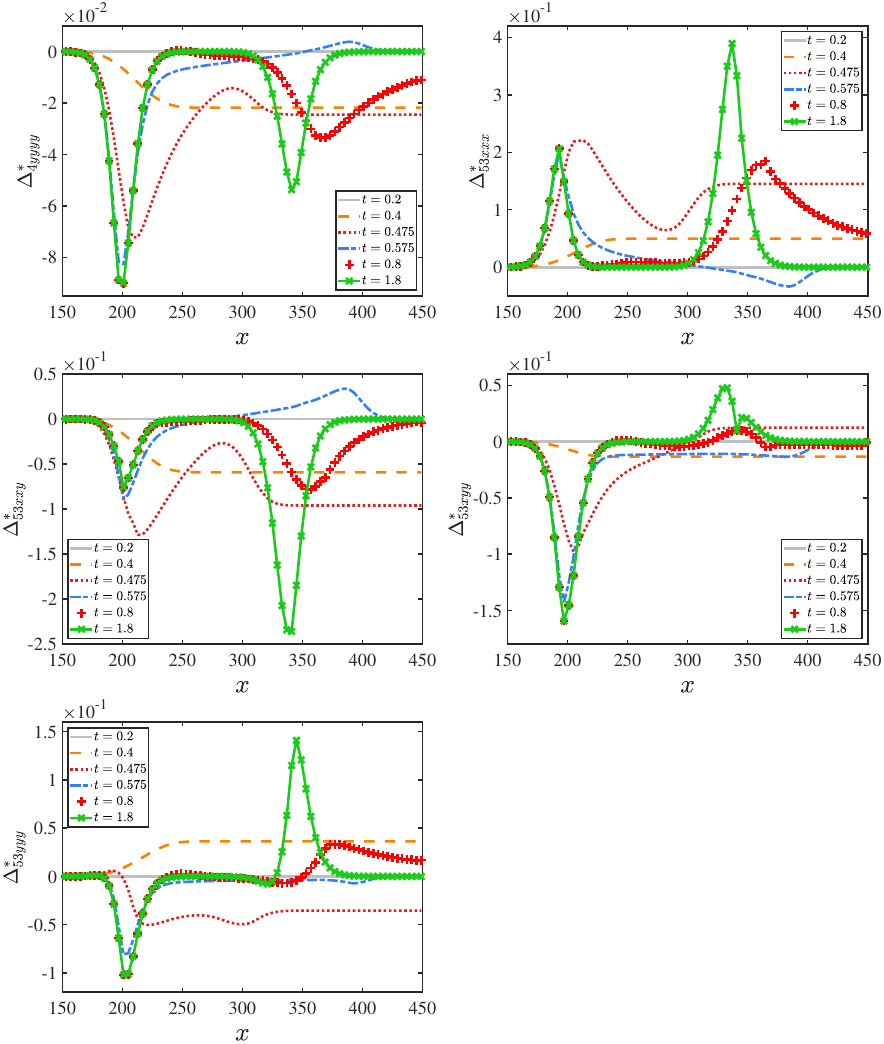, width=\textwidth}
        \label{fig:sub3}
    \end{subfigure}
\caption{Distributions of typical nonequilibrium quantities $\bm{\Delta}%
_{2}^*$, $\bm{\Delta}_{3,1}^*$, $\bm{\Delta}_{3}^*$, $\bm{\Delta}_{4,2}^*$, $
\bm{\Delta}_{4}^*$, and $\bm{\Delta}_{5,3}^*$ along the line $y=40\Delta y$
at various instants.}
\label{RRT3}
\end{figure*}

It should be noted that the total nonequilibrium strength shown in figures \ref{RR} and
\ref{Ma2-5} is presented in a highly condensed form to represent the nonequilibrium state.
The TNE characteristic quantities $\bm{\Delta}_{2}^*$,
$\bm{\Delta}_{3,1}^*$, $\bm{\Delta}_{3}^*$, $\bm{\Delta}_{4,2}^*$, $\bm{\Delta}%
_{4}^*$, and $\bm{\Delta}_{5,3}^*$ reflect the system's deviation from
equilibrium from their respective perspectives, as illustrated in figure \ref{RRT3}.
These measures provide detailed nonequilibrium information at various
temporal and spatial scales embedded in the flow field, which can be used to
identify shock wave types, track their positions, and characterize their
features.

From figure \ref{RRT3}, the following common features can be observed:

(I) All nonequilibrium quantities, $\bm{\Delta}_{2}^*$, $\bm{\Delta}_{3,1}^*$%
, $\bm{\Delta}_{3}^*$, $\bm{\Delta}_{4,2}^*$, $\bm{\Delta}_{4}^*$, and $%
\bm{\Delta}_{5,3}^*$, exhibit peak values near the shock wave interface and
decay to zero away from it. This is due to the large gradients in macroscopic
quantities and strong nonequilibrium driving forces near the shock
interface, causing a significant deviation from equilibrium.

(II) At the location of the incident shock wave, the peak positions of the
nonequilibrium quantities shift downward over time. After the shock reaches
the lower wall ($t > 0.575$), under the influence of the solid wall, the peak
positions of both the incident and reflected shock waves
continue to shift leftward, while their amplitudes increase (except for $%
\Delta_{2xx}^*$).

(III) At lower Mach numbers ($Ma = 2.2$), the number of nonequilibrium effects dominant at the incident shock is approximately equal to those at the reflected shock. The relative dominance of different types of nonequilibrium effects within the same physical process varies, reflecting the complexity of the system and highlighting the need for multi-perspective studies to fully understand its behavior.

(IV) For each nonequilibrium component, the first peak's appearance on the
left (right) side corresponds to the time when the incident (reflected)
shock wave moves downward (upward) to the current horizontal line. By
observing the timing and position of the first peak at various heights, the
formation of the incident and reflected shock waves can be tracked.

(V) Compared to $t = 1.8$, the deviation from equilibrium region
is broader but weaker for $t < 0.8$, suggesting a wider but less
intense curved reflected shock wave interface. However, at $t = 0.8$, most
nonequilibrium components at the curved reflected shock wave location
exhibit a significant amplitude increase, as the curved shock wave crosses the
horizontal line $y = 40\Delta y$. This characteristic can be used to track
the position of the curved shock wave and identify its transition to a
straight shock wave. For example, at $t = 0.8$, the position $x = 300$
corresponds to $\Delta_{2xx}^*$, aligning with the intersection of the
curved shock wave and the line $y = 40\Delta y$, where all nonequilibrium
quantities display this feature.

(VI) At $t = 0.575$ and $t = 0.8$, the peak positions and amplitudes of the
nonequilibrium components at the incident shock wave location are nearly
identical. This provides strong evidence that the incident shock wave is
about to compress and straighten, signaling its transition from a curved to
an oblique shock wave. Similarly, at $t = 1.3$ and $t = 1.8$, the peak
positions and amplitudes of the nonequilibrium components at the reflected shock wave are also
very close, providing clear evidence that the reflected shock wave is about to compress and straighten, signaling the transition from a curved to an oblique shock wave.

In addition to the common features outlined above, the differences among
various nonequilibrium components are discussed as follows:

(I) The driving mechanisms of different nonequilibrium components are
distinct, each describing the system¡¯s deviation from
equilibrium from a unique perspective. These components are independent, non-interchangeable, and complementary, collectively providing an intuitive representation of the system's complex behavior.

(II) As time evolves, the peak value of $\Delta_{2xx}^*$ at the location of
the incident shock wave first decreases negatively, then increases positively. The
shock interface initially widens and then gradually narrows, becoming
progressively compressed.

(III) For $t > 0.575$, as the incident shock wave moves downward, the
nonequilibrium interface, observed from the $\Delta_{2xy}^*$ perspective,
gradually broadens under the influence of the curved shock wave.
Furthermore, the shear viscosity effect intensifies, causing the shock wave
to further curve and dissipate as it moves downward, resulting in a more
extensive deviating-from-equilibrium region.

(IV) As the incident shock wave moves downward and reflects off the lower
wall, energy is gradually transferred to the reflected shock wave,
significant increasing $\Delta_{2xy}^*$ and $\Delta_{2yy}^*$ at the
reflected shock wave. Before the peak at the reflected shock wave (at $t =
1.8$), the curves of $\Delta_{2xx}^*$ and $\Delta_{2yy}^*$ exhibit symmetry.

(V) At $t = 1.8$, $\Delta_{3,1x}^*$ reaches its maximum at the reflected
shock wave, indicating an intense interaction between the incident and
reflected shock waves, with significant heat exchange. In addition, at this
moment, the heat flux in the $y$-direction, $\Delta_{3,1y}^*$, shows a
positive peak at the reflected shock wave, exceeding $\Delta_{3,1x}^*$,
indicating greater heat transfer in the $y$-direction. This difference in
peak values reveals the anisotropy of heat transfer during the interaction
between the incident and reflected shock waves.

(VI) When $t > 0.575$, the viscous stress fluxes $\Delta_{3xxy}^*$, $%
\Delta_{3xyy}^*$, and $\Delta_{3yyy}^*$ gradually reach their peak values at
the curved incident/reflected shock waves. As the shock waves straighten,
their peak values increase further, indicating a stronger molecular motion
within the shock. When $0.475 \leq t \leq 0.8$, $\Delta_{3xxx}^*$ reaches an
equilibrium at the center of the shock wave, then deviates oppositely to
reach a peak before returning to equilibrium further from the shock
interface. These observations highlight the directional differences in the
viscous stress flux $\bm{\Delta}_3^*$, reflecting the anisotropic nature of
the fluid¡¯s viscous response in different directions.

(VII) At $t = 0.575$, the flux of heat flux $\Delta_{4,2xy}^*$
reaches its peak at the center of the curved incident shock.
In contrast, $\Delta_{4,2xx}^*$ and $\Delta_{4,2yy}^*$ reach equilibrium at the center of
the curved incident shock, with peaks on either side.
When $t \geq 0.8$, at
the center of the reflected shock, the peaks of $\Delta_{4,2xx}^*$ and $%
\Delta_{4,2xy}^*$ shift leftward over time, with their intensities gradually
increasing. These findings provide key instantaneous characteristics for understanding the
thermodynamic flux dynamics during regular shock wave reflection.

(VIII) For $t \geq 0.475$, peaks of $\Delta_{4xxxx}^*$, $\Delta_{4xxxy}^*$, $%
\Delta_{4xxyy}^* $, $\Delta_{4xyyy}^*$, and $\Delta_{4yyyy}^*$ are observed
at the curved incident shock location. Additionally, $\Delta_{4xxxx}^*$, $%
\Delta_{4xxxy}^* $, and $\Delta_{4yyyy}^*$ reach positive peaks, while the
remaining components exhibit negative values. Between $0.475 \leq t \leq 0.575$%
, these peaks shift leftward within the curved reflected shock region. These
findings highlight temporal evolution characteristics of the anisotropic
components of the viscous stress flux $\Delta_4^*$ at the location of the
curved incident shock. For $t > 0.5$ the peaks of $%
\Delta_{4xxxx}^*$, $\Delta_{4xxyy}^*$, $\Delta_{4xyyy}^*$, and $%
\Delta_{4yyyy}^*$ at the center of the curved reflected shock
gradually increase over time, indicating intensified heat flux in specific directions
during shock reflection.

(IX) For $t \geq 0.475$, peaks of $\Delta_{5,3xxx}^*$, $\Delta_{5,3xxy}^*$, $%
\Delta_{5,3xyy}^*$, and $\Delta_{5,3yyy}^*$ are observed at the curved
incident shock location. These peaks are particularly significant across all
observation times. Notably, the peak value of $\Delta_{5,3xxx}^*$ is
positive, while the peak values of $\Delta_{5,3xxy}^*$, $\Delta_{5,3xyy}^*$,
and $\Delta_{5,3yyy}^*$ are negative. This phenomenon also reveals the
anisotropic spatial characteristics of the heat flux components. For $t >
0.8 $, the peaks of $\Delta_{5,3xxx}^*$, $\Delta_{5,3xxy}^*$, $\Delta_{5,3xyy}^*$,
and $\Delta_{5,3yyy}^*$ at the curved reflected shock location gradually
increase. This indicates that, as time progresses, the nonequilibrium heat
flux dynamics in the curved reflected shock region continue to intensify,
reflecting the complex evolution of the fluid structure during shock
interaction.

\subsection{Entropy production mechanism}

\begin{figure*}
\centering{%
\epsfig{file=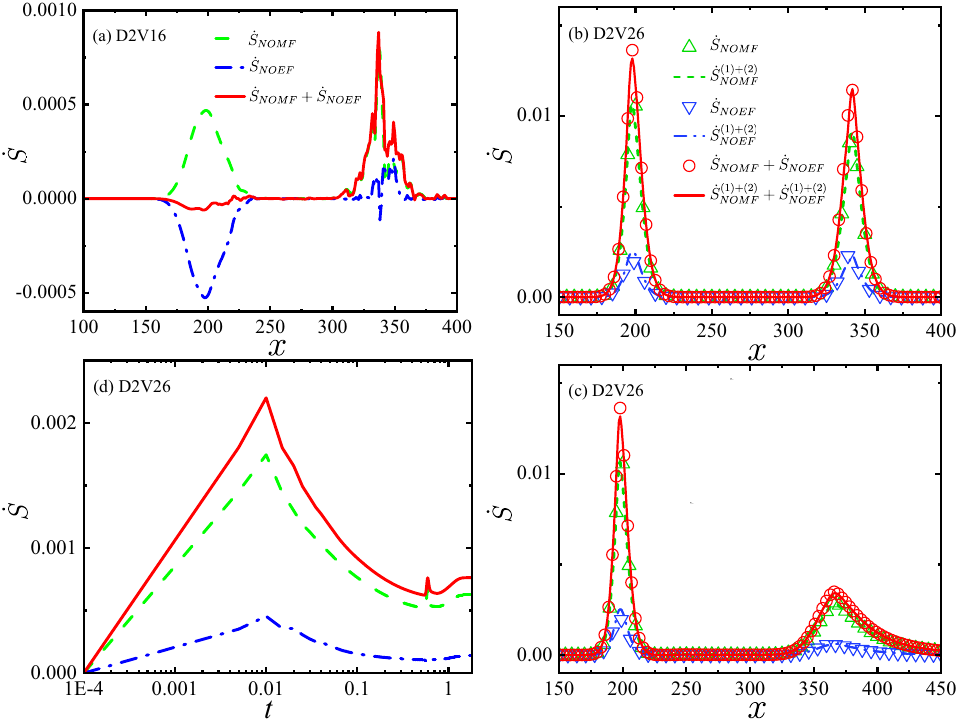,
			width=0.99\textwidth}}
\caption{Entropy production rates obtained from the first-order DBM (a) and
second-order DBM along the line $y = 40\Delta y$ at $t=1.8$, where the lines
in (b) corresponding to theoretical solutions. (c) Entropy production rates
along the same line at $t=0.8$. (d) Evolution of entropy production rates
during the regular shock wave reflection.}
\label{CompareEntropy}
\end{figure*}

Figure \ref{CompareEntropy}(c) shows the entropy production rates from
viscous stress, heat flux, and their sum along the line $y = 40 \Delta y$ at
$t = 0.8$. The symbols represent DBM simulation results, while the lines
show the corresponding second-order analytical solutions. Comparing with
figure \ref{RRT3}, it is clear that the entropy production rate is highly
correlated with the nonequilibrium effects.

The entropy production rate
peaks at $x = 198$ and $342$, where both viscous stress and heat flux reach
their maxima, exactly corresponding to the positions of incident
and curved reflected shock waves. Therefore, by observing the peak positions
and shapes of the entropy production rate curve, the evolution speed
and morphological distribution of the shock waves can be dynamically characterized from a more
fine view. In comparison with figure \ref{CompareEntropy}(c) at $t = 1.8$,
the nonzero region of entropy production at the reflected shock wave is
wider but less intense, due to the broad, uncompressed curved shock wave
interface. As the reflected shock wave interacts with the wall, the curved
shock wave is compressed and straightened, increasing the intensity of
entropy production and shifting the peak position leftward. Throughout the
regular shock wave reflection process, the contribution of heat flux to
entropy production is much smaller than that of viscous stress, especially
at the curved shock wave, indicating that the process is predominantly by
viscous stress.

Figure \ref{CompareEntropy}(d) illustrates the evolution of the three types
of entropy production over time. The entropy production rate
initially increases, then decreases, increases again, and finally stabilizes. At
$t = 0.01$, the entropy production rate rapidly reaches
its maximum due to the nonequilibrium driving force caused by the initial macroscopic gradient.
At this point, the system has not yet undergone dissipation. The
shock interface is narrow, and the entropy production rate due to stress reaches
its peak earlier than that due to heat flux. This suggests that viscous
stress, driven by sustained incoming flow from the left and upper
boundaries, is the dominant driving force of the system. At $t = 0.59$, a local
peak in entropy production arises due to the formation of the curved
incident shock wave, which enlarges the nonequilibrium region. Afterward, as
the incident shock wave interacts with the lower wall and the curved
reflected shock wave forms, entropy production slowly increases,
until $t = 1.35$, when it stabilizes.

\begin{figure}
\centering{%
\epsfig{file=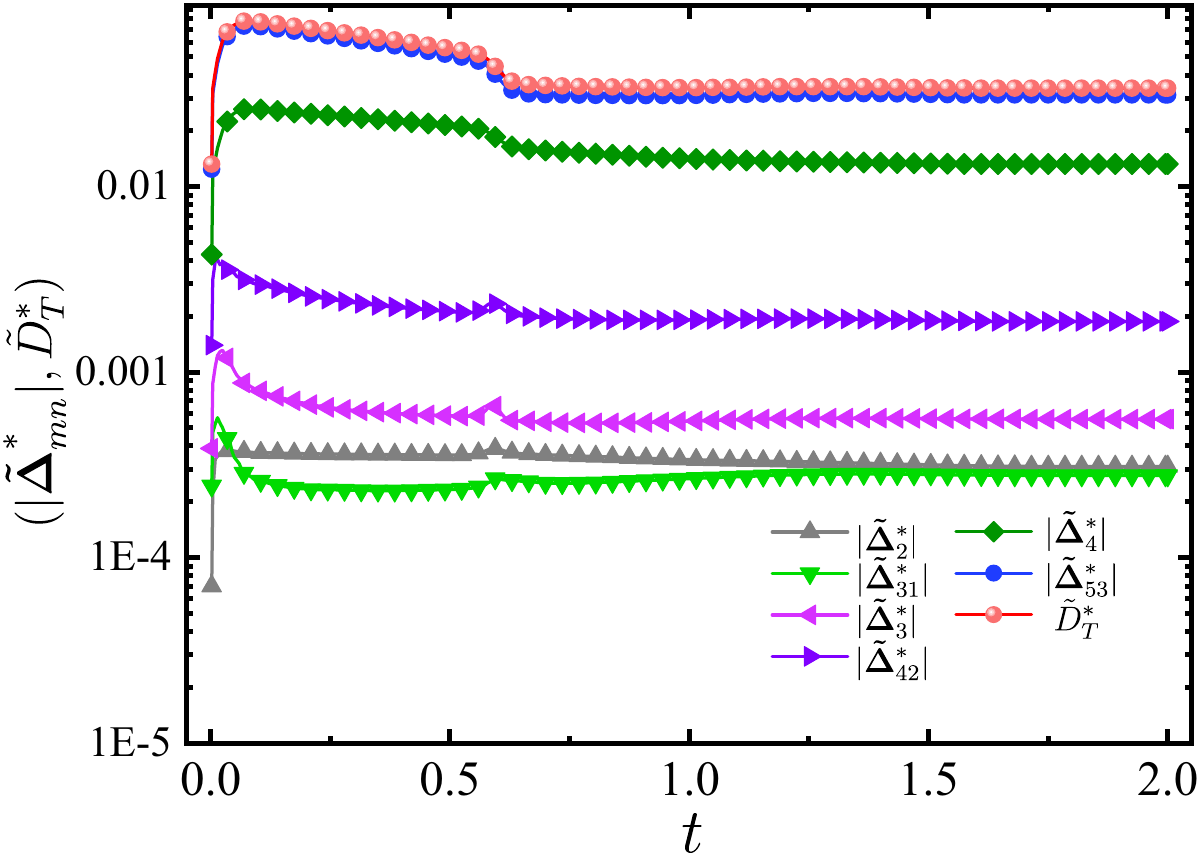,
			width=0.6\textwidth}}
\caption{Evolution of dimensionless nonequilibrium intensities from various
perspectives.}
\label{D-Dim}
\end{figure}

Figure \ref{D-Dim} shows the temporal evolution of dimensionless
nonequilibrium intensities. The intensities of various nonequilibrium
quantities span several orders of magnitude. Starting from an initial
nonequilibrium state, $|\bm{\Delta}_{m,n}^*|$ first shows a sharp increase,
followed by a gradual approach toward thermodynamic equilibrium. This
behavior results from the strong correlation between TNE effects and
interface dynamics, which are characterized by varying area and morphological
complexity.

Higher-order TNE dynamic modes play a more significant role compared to
their lower-order counterparts, and dominate the total nonequilibrium
intensity. Additionally, the relative strength of dynamic modes varies across
different stages. For example, $|\bm{\Delta}{2}^*|$ is initially weaker
than $|\bm{\Delta}{3,1}^*|$, but this relationship reverses at later stages.
Overall, their evolution trends closely align with that of entropy
production rates. At $t = 0.6$, a slight jump in $|\bm{\Delta}{3,1}^*|$ and $|%
\bm{\Delta}{4,2}^*|$ marks the moment when the incident shock wave reaches
the bottom wall. This change serves as an indicator for determining the
time of shock wave reflection.

In fact, the entropy production rate and TNE intensity, analyzed from
different perspectives, provide a comprehensive view of the system¡¯s
nonequilibrium characteristics. These detailed dynamics complement the
synthetic TNE strength $D_T^*$ and macroscopic conserved quantities
typically considered in traditional fluid mechanics.

\subsection{Effects of Mach number on thermodynamic nonequilibrium effects}

In practical engineering applications, the range of Mach numbers involved is
quite broad. A higher Mach number corresponds to a greater compression
effect and higher pressure ratio, density ratio, and temperature jump,
resulting in stronger interaction between the shock wave and the medium. To
explore the influence of Mach number on nonequilibrium intensity during
oblique shock wave reflection, simulations were performed for Mach numbers
ranging from $2.1$ to $5.1$, with increments of $0.3$ and an incident angle
of $29^\circ$.

\begin{figure}
\centering{%
\epsfig{file=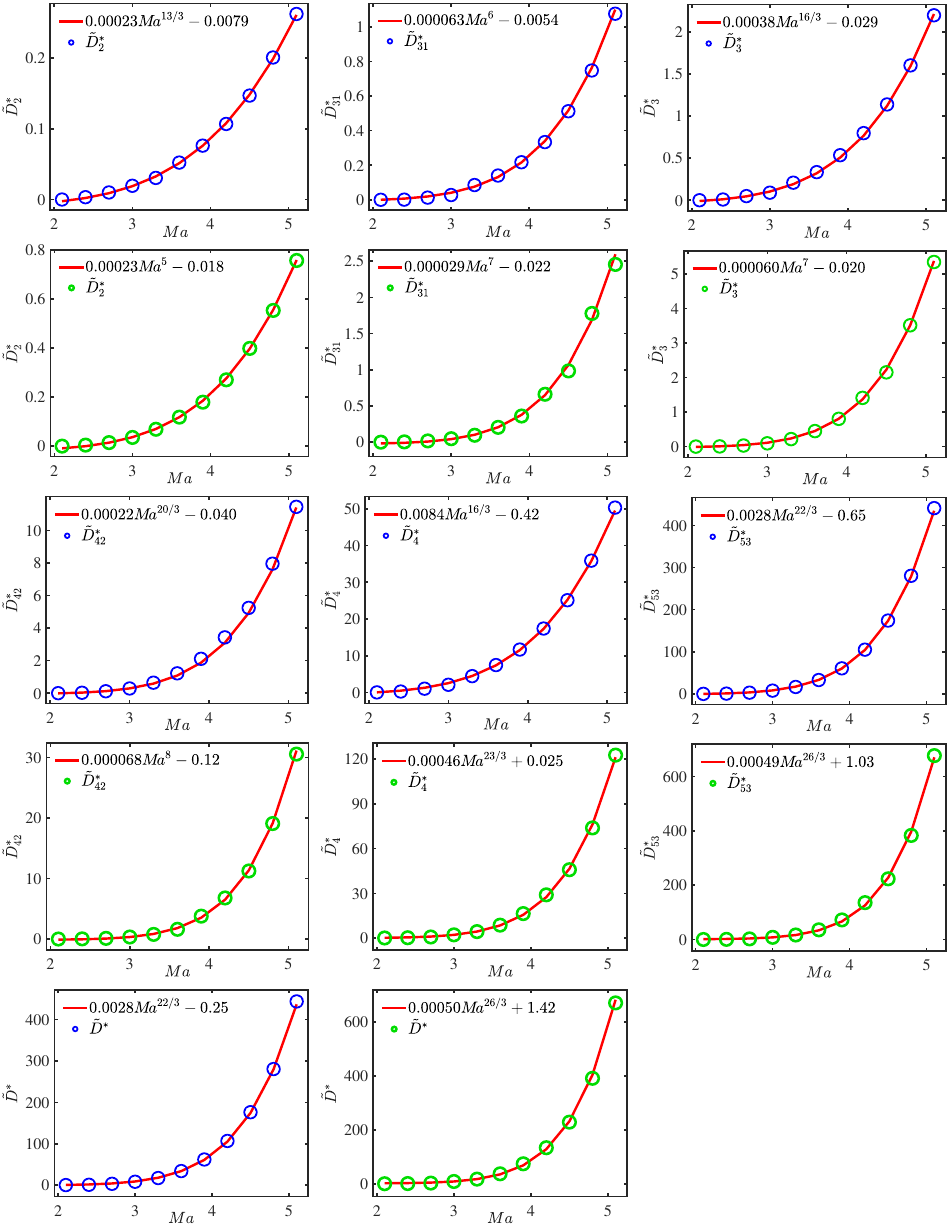,
			width=0.99\textwidth}}
\caption{Effects of Mach number on the peak values of various nonequilibrium
measures. 
The blue circles represent the DBM simulation results at the incident shock, while the green circles correspond to those at the reflected shock. The red solid lines show the nonlinear fits to the simulation data.}
\label{Ma-D}
\end{figure}

Figure \ref{Ma-D} illustrates the effects of Mach number on the peak values
of various nonequilibrium measures at $t = 0.8$ and $y = 10 \Delta y$. The
left column shows results for the incident shock, and the right column
corresponds to the reflected shock. The circles and solid lines in each
panel represent the DBM simulation results and the nonlinear fitting
results, respectively. From figure \ref{Ma-D}, it is evident that:

(I) The nonequilibrium intensity increases significantly with the order $m$
of the nonequilibrium quantities.

(II) The Mach number $Ma$ exponentially amplifies the nonequilibrium
intensity $\tilde{D}_{m,n}^*$.

(III) The growth exponent of the Mach number $\alpha$ for higher-order
nonequilibrium quantities is substantially larger than that for lower-order
quantities. For instance:
\begin{itemize}
\item $\tilde{D}_2: \sim Ma^{13/3}$ (left column) $\sim Ma^{5}$ (right
column),

\item $\tilde{D}_{5,3}: \sim Ma^{22/3}$ (left column) $\sim Ma^{26/3}$ (right
column).
\end{itemize}
This suggests that higher-order quantities are more sensitive to variations
in the Mach number.

(IV) The peak values of nonequilibrium intensity at the reflected shock are
consistently higher than those at the incident shock. This
difference arises from the interaction between the reflected shock and the
wall, which generates stronger local nonequilibrium effects. Additionally,
this highlights the greater sensitivity of the nonequilibrium intensity at the
reflected shock to variations in the Mach number.

(V) The intercept (constant term) of the fitted curves is nearly zero for
the incident shock but much larger for the reflected shock. This behavior is
attributed to the more complex structure and stronger local macroscopic
gradients at the reflected shock.

\begin{figure}
\centering{%
\epsfig{file=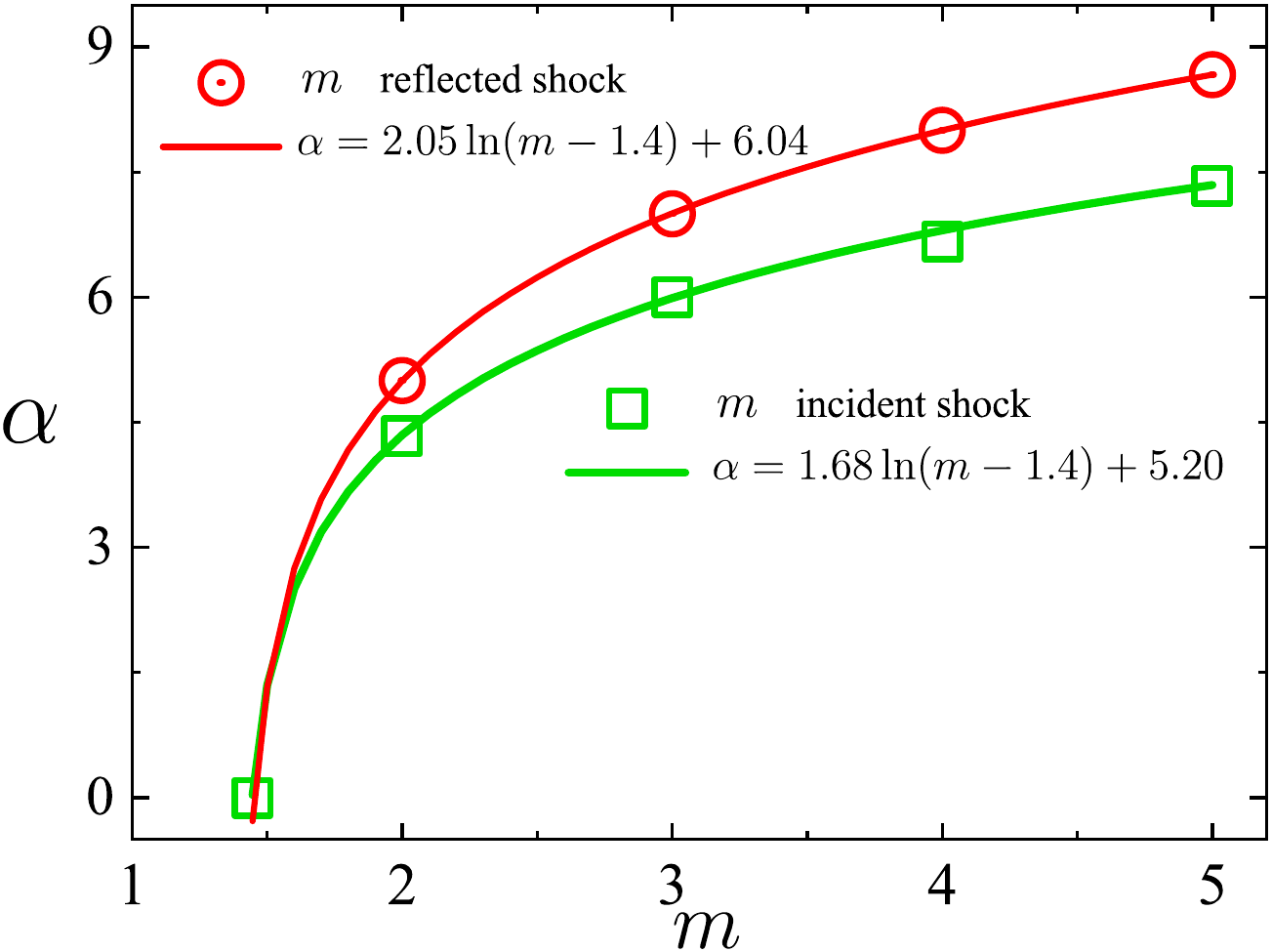,
			width=0.55\textwidth}}
\caption{The correlations between the Mach number growth exponent $\protect%
\alpha$ and the order of nonequilibrium quantities $m$ at both the incident and
reflected shocks.}
\label{log}
\end{figure}

To further clarify the trends disscussed above, figure \ref{log} presents the
correlations between the Mach number growth exponent $\alpha$ and the order
of nonequilibrium quantities $m$ for both the incident and reflected shocks.
Key observations include: (1) For both incident and reflected shocks, the growth
exponent $\alpha$ increases logarithmically with the order $m$ of the
nonequilibrium quantities: $\alpha \sim \ln(m - c)$, where $c$ is a
constant. This indicates that higher-order nonequilibrium quantities are more
sensitive to variations in the Mach number. As $m$ increases, the growth rate of
$\alpha$ slows but still shows a progressive enhancement.
(2) The curve for the reflected shock consistently lies above the curve for the incident shock
and has a larger intercept. For the same order of nonequilibrium quantity $m$%
, the growth exponent $\alpha$ is considerably higher at the reflected shock
than at the incident shock. This underscores the stronger nonequilibrium effects at the reflected
shock.

\section{Discussion, conclusions, and recommendations}
\label{Conclusions}

The challenges in simulating supersonic flow can be categorized into two
types: one arises from the equation solvers, such as discrete schemes, and
the other from the physical model itself.
The limitations of the physical
model cannot be overcome by enhancing algorithmic accuracy and efficiency.
Early studies predominantly relied on the Navier-Stokes (NS) equations,
which are based on the quasi-continuum assumption and the near-equilibrium
approximation. However, mesoscale structures, such as shock waves, are
small-scale structures, and compared to the shock thickness, the mean
molecular spacing is no longer a negligible small quantity, leading to
increased discreteness.
The physical quantities across the shock change
abruptly, signifying a rapid process. As these quantities vary, the system
lacks sufficient time to return to thermodynamic equilibrium, resulting in a
higher degree of nonequilibrium.
The quasi-continuum model fails to capture
discrete effects, such as rarefaction, while the near-equilibrium model
is inadequate for describing strong non-equilibrium behaviors.
Furthermore, the
lack of advanced techniques for analyzing complex physical fields results in
significant data resource waste.
As discretization and rarefaction increase,
the description of the system state demands more state variables, and the
characterization of effects necessitates additional variables.
This is an inevitable requirement to accurately describe physical mechanisms as discretization
and rarefaction increase.
The primary technical challenge is determining which physical variables need to be added. A
typical
feature of supersonic flow is that even a small change can trigger widespread
effects. Previous insights, based on slow variables, conserved quantities,
and the Knudsen number defined from a single perspective, have been incomplete or inaccurate.

In response to the these issues, this paper develops a more effective
approach for constructing physical models and a more refined technique for
analyzing complex physical fields-the discrete Boltzmann method (DBM), to
study the multiscale, nonequilibrium, and complex behavior of supersonic
flows.
The key conclusions are summarized as follows: (1) A Burnett-level DBM model,
tailored for supersonic flows, is developed based on the Shakhov-BGK
framework. It provides higher-order analytical expressions for typical
thermodynamic nonequilibrium effects, forming a constitutive foundation for
improving traditional macroscopic hydrodynamic models.
(2) Criteria forvalidating the DBM model are established by comparing numerical solutions
 with analytical results for nonequilibrium effects. (3) The multiscale DBM
is applied to analyze nonequilibrium characteristics and entropy production
mechanisms in shock wave reflections. Compared to Navier-Stokes-level DBM,
the Burnett-level DBM offers: (a) More accurate representations of viscous
stress and heat flux, i.e., dissipative mechanisms; (b) Consistency with the
second law of thermodynamics by ensuring non-negative entropy production;
(c) Improved numerical stability. (d) Near the interfaces of incident and
reflected shock waves, strong nonequilibrium driving forces result in
prominent nonequilibrium effects. The distribution and evolution of these
effects in shock reflections suggest that nonequilibrium quantities are
critical indicators for tracking complex flow processes. This is
particularly valuable for understanding shock interface dynamics. (e) In
intermediate states of shock reflections, the bent reflected shock and incident
shock interfaces are wider and exhibit lower nonequilibrium intensities than
in the final state. This observation can be used to track the position and
straightening process of the bent shock.
(f) The Mach number enhances
nonequilibrium intensities in a power-law manner. The power-law exponent and
the order of nonequilibrium dynamic modes follow a logarithmic relationship.
This study provides kinetic insights into the modeling and analysis of
supersonic flows, providing additional perspectives for further exploration
of nonequilibrium and multiscale phenomena in complex flow systems.

Future research should focus on the following directions: (1) Model
extensions: Further expand DBM models to accommodate more complex physical
scenarios, such as chemically reactive or multiphase supersonic flows. (2)
Nonequilibrium diagnostics and regulations: Develop diagnostic and control
techniques based on nonequilibrium quantities and corresponding analytical
solutions for achieving engineering optimization of complex flow processes.


\section*{Acknowledgements}

We acknowledge support from the National Natural Science Foundation of China
(Grant Nos. 11875001 and 12172061), Hebei Outstanding Youth Science
Foundation (Grant No. A2023409003), the Central Guidance on Local Science
and Technology Development Fund of Hebei Province (Grant No. 226Z7601G),
 Foundation of National Key Laboratory
of Shock Wave and Detonation Physics (Grant No.
JCKYS2023212003), and the opening project of State Key Laboratory of
Explosion Science and Technology (Beijing Institute of Technology)
(Grant No. KFJJ23-02M), the Foundation of the National Key Laboratory of Computational Physics (Grant No. SYSQN2024-10), the High-level Talents Research Start-up Grant from Guangxi University (Grant No. ZX01080031224009).

\section*{Declaration of interests}

The authors report no conflict of interest.


\appendix

\section{Second-order analytical expressions for viscous stress and heat flux%
}

\label{appendix1}

The contributions of the second-order deviation of the distribution function
$f^{(2)}$ to viscous stress and heat flux are detailed as follows::
\begin{equation}
\Delta _{2xx}^{\ast (2)}=2n_{2}^{-1}{\tau ^{2}}\left\{
\begin{array}{l}
n_{2}^{-1}\rho RT\left[
\begin{array}{l}
{n_{-2}}{n_{1}}{\left( {{\partial _{x}}{u_{x}}}\right) ^{2}}+{n_{1}}{n_{2}}{%
\left( {{\partial _{y}}{u_{x}}}\right) ^{2}} \\
-4n{\partial _{x}}{u_{x}}{\partial _{y}}{u_{y}}-{n_{2}}{\left( {{\partial
_{x}}{u_{y}}}\right) ^{2}}-{n_{-2}}{\left( {{\partial _{y}}{u_{y}}}\right)
^{2}}%
\end{array}%
\right] \\
-{R^{2}}{T^{2}}({n_{1}}\frac{{{\partial ^{2}}\rho }}{{\partial {x^{2}}}}-%
\frac{{{\partial ^{2}}\rho }}{{\partial {y^{2}}}}) \\
+\frac{{{R^{2}}{T^{2}}}}{\rho }\left[ {{n_{1}}{{\left( {{\partial _{x}}\rho }%
\right) }^{2}}-{{\left( {{\partial _{y}}\rho }\right) }^{2}}}\right] +\frac{{%
\rho {R^{2}}}}{{\Pr }}\left[ {{n_{1}}{{\left( {{\partial _{x}}T}\right) }^{2}%
}-{{\left( {{\partial _{y}}T}\right) }^{2}}}\right] \\
+\rho {R^{2}}T(\frac{1}{{\Pr }}-1)({n_{1}}\frac{{{\partial ^{2}}T}}{{%
\partial {x^{2}}}}-\frac{{{\partial ^{2}}T}}{{\partial {y^{2}}}})+{R^{2}}T(%
\frac{1}{{\Pr }}-1)({n_{1}}{\partial _{x}}\rho {\partial _{x}}T-{\partial
_{y}}\rho {\partial _{y}}T)%
\end{array}%
\right\}  \label{vis-2nd}
\end{equation}%
\begin{equation}
\Delta _{2xy}^{\ast (2)}=2{\tau ^{2}}\left\{
\begin{array}{l}
\rho RTn_{2}^{-1}(-2{\partial _{x}}{u_{x}}{\partial _{y}}{u_{x}}+n{\partial
_{x}}{u_{x}}{\partial _{x}}{u_{y}}+n{\partial _{y}}{u_{x}}{\partial _{y}}{%
u_{y}}-2{\partial _{x}}{u_{y}}{\partial _{y}}{u_{y}}) \\
-{R^{2}}{T^{2}}\frac{{{\partial ^{2}}\rho }}{{\partial x\partial y}}+\frac{{{%
R^{2}}{T^{2}}}}{\rho }{\partial _{x}}\rho {\partial _{y}}\rho +\frac{{\rho {%
R^{2}}}}{{Pr}}{\partial _{x}}T{\partial _{y}}T+\rho {R^{2}}T(\frac{1}{{\Pr }}%
-1)\frac{{{\partial ^{2}}T}}{{\partial x\partial y}} \\
+\frac{{{R^{2}}T}}{2}(\frac{1}{{\Pr }}-1)\left( {{\partial _{x}}\rho {%
\partial _{y}}T+{\partial _{y}}\rho {\partial _{x}}T}\right)%
\end{array}%
\right\}
\end{equation}%
\begin{equation}
\Delta _{2yy}^{\ast (2)}=2n_{2}^{-1}{\tau ^{2}}\left\{
\begin{array}{l}
n_{2}^{-1}\rho RT\left[
\begin{array}{l}
{n_{-2}}{n_{1}}{\left( {{\partial _{y}}{u_{y}}}\right) ^{2}}+{n_{1}}{n_{2}}{%
\left( {{\partial _{x}}{u_{y}}}\right) ^{2}} \\
-4n{\partial _{x}}{u_{x}}{\partial _{y}}{u_{y}}-{n_{2}}{\left( {{\partial
_{y}}{u_{x}}}\right) ^{2}}-{n_{-2}}{\left( {{\partial _{x}}{u_{x}}}\right)
^{2}}%
\end{array}%
\right] \\
-{R^{2}}{T^{2}}({n_{1}}\frac{{{\partial ^{2}}\rho }}{{\partial {y^{2}}}}-%
\frac{{{\partial ^{2}}\rho }}{{\partial {x^{2}}}})+\frac{{{R^{2}}{T^{2}}}}{%
\rho }\left[ {{n_{1}}{{\left( {{\partial _{y}}\rho }\right) }^{2}}-{{\left( {%
{\partial _{x}}\rho }\right) }^{2}}}\right] \\
+\frac{{\rho {R^{2}}}}{{\Pr }}\left[ {{n_{1}}{{\left( {{\partial _{y}}T}%
\right) }^{2}}-{{\left( {{\partial _{x}}T}\right) }^{2}}}\right] +\rho {R^{2}%
}T(\frac{1}{{\Pr }}-1)({n_{1}}\frac{{{\partial ^{2}}T}}{{\partial {y^{2}}}}-%
\frac{{{\partial ^{2}}T}}{{\partial {x^{2}}}}) \\
+{R^{2}}T(\frac{1}{{\Pr }}-1)({n_{1}}{\partial _{y}}\rho {\partial _{y}}T-{%
\partial _{x}}\rho {\partial _{x}}T)%
\end{array}%
\right\}  \label{heat-2nd}
\end{equation}%
\begin{equation}
\Delta _{3,1x}^{\ast (2)}=n_{2}^{-1}\frac{{{\tau ^{2}}\rho {R^{2}}T}}{{\Pr^{2}%
}}\left\{
\begin{array}{l}
T[(2{n_{1}}\Pr -{n_{4}})\frac{{{\partial ^{2}}{u_{x}}}}{{\partial {x^{2}}}}%
+(n\Pr -{n_{4}})\frac{{{\partial ^{2}}{u_{y}}}}{{\partial x\partial y}}+{%
n_{2}}\Pr \frac{{{\partial ^{2}}{u_{x}}}}{{\partial {y^{2}}}}] \\
+({n_{1}}{n_{6}}\Pr +{n_{-2}}){\partial _{x}}{u_{x}}{\partial _{x}}T+\frac{{{%
n_{2}}{n_{6}}(\Pr +1)}}{2}{\partial _{y}}{u_{x}}{\partial _{y}}T \\
-{n_{6}}(\Pr +1){\partial _{y}}{u_{y}}{\partial _{x}}T+(\frac{{{n_{2}}{n_{6}}%
\Pr -{n_{2}}^{2}}}{2}){\partial _{x}}{u_{y}}{\partial _{y}}T%
\end{array}%
\right\}  \label{heat-2nd}
\end{equation}%
\begin{equation}
\Delta _{3,1y}^{\ast (2)}=n_{2}^{-1}\frac{{{\tau ^{2}}\rho {R^{2}}T}}{{\Pr^{2}%
}}\left\{
\begin{array}{l}
T[(2{n_{1}}\Pr -{n_{4}})\frac{{{\partial ^{2}}{u_{y}}}}{{\partial {y^{2}}}}%
+(n\Pr -{n_{4}})\frac{{{\partial ^{2}}{u_{x}}}}{{\partial x\partial y}}+{%
n_{2}}\Pr \frac{{{\partial ^{2}}{u_{y}}}}{{\partial {x^{2}}}}] \\
+({n_{1}}{n_{6}}\Pr +{n_{-2}}){\partial _{y}}{u_{y}}{\partial _{y}}T+\frac{{{%
n_{2}}{n_{6}}(\Pr +1)}}{2}{\partial _{x}}{u_{y}}{\partial _{x}}T \\
-{n_{6}}(\Pr +1){\partial _{x}}{u_{x}}{\partial _{y}}T+(\frac{{{n_{2}}{n_{6}}%
\Pr -{n_{2}}^{2}}}{2}){\partial _{y}}{u_{x}}{\partial _{x}}T%
\end{array}%
\right\}
\end{equation}%
Here $n_{a}=n+a$, for example $n_{1}=n+1$.

The second-order analytical expressions indicate that viscous stress depends
not only on velocity gradients but also on additional terms, including:

(1) Squared velocity gradient terms, such as $(\partial_{\alpha}
u_{\alpha})^2$ and $(\partial_{\alpha} u_{\beta})^2$, capture the nonlinear
effects of velocity variations. These terms represent the intensity of local
variations rather than just the direction or rate of change. Such intense
variations can lead to local stress concentrations. For example, $%
(\partial_{\alpha} u_{\alpha})^2$ may cause local compression or
rarefaction, while $(\partial_{\alpha} u_{\beta})^2$ can induce local shear.

(2) The squared term of density gradients, $(\partial_{\alpha} \rho)^2$,
reflects the intensity of density variations and affects local compressive
stress and flow stability

(3) The second-order derivatives of density, $\frac{\partial^2 \rho}{%
\partial {\alpha}^2}$, represent the curvature of density in $\alpha$
direction. This curvature reflects the degree of bending in the density
distribution and the compressive behavior of the fluid in specific regions.
Such changes in curvature can lead to local volume responses and variations
in stress within the fluid. Both terms (2) and (3) highlight the impact of
density inhomogeneity on viscous stress, mediated by the compressibility of
the fluid.

(4) The squared temperature gradient term, $(\partial_{\alpha} T)^2$,
reflects the intensity of temperature gradients and mainly affects thermal
convection. This, in turn, influences viscous stress by altering both the
strength and mode of convection.

(5) The second-order derivatives of temperature, $\frac{\partial^2 T}{%
\partial {\alpha}^2}$, reflect the curvature of temperature and primarily
affect the rate of thermal diffusion. Consequently, they influence viscous
stress by affecting heat diffusion within the fluid.

(6) Coupled viscous stress terms, such as $\partial_{\alpha} \rho
\partial_{\alpha} T $, represent the interaction between density and
temperature gradients.

The term $\Delta _{2xy}^{ * (2)}$ includes additional coupling and
cross-derivative terms. The term $\Delta _{2yy}^{ * (2)}$ contains the same
additional terms as $\Delta _{2xx}^{ * (2)}$.

Heat flux depends not only on temperature gradients but also on additional
terms, including:

(1) The second-order derivatives of velocity, $\frac{\partial^2 u_{\alpha}}{%
\partial \beta^2}$, reflect contribution of fluid diffusion to heat flux.

(2) Coupled heat flux terms, such as $\partial_{\beta} u_{\beta}
\partial_{\alpha} T $, $\partial_{\beta} u_{\alpha} \partial_{\beta} T $,
and $\partial_{\alpha} u_{\beta} \partial_{\beta} T $, reflect how velocity
gradients regulate thermal flux.

\section{Independent kinetic moments required to fully characterize first-
and second-order viscous stress and heat flux}

\label{appendix2}

To fully characterize first- and second-order viscous stress and heat flux,
the Shakhov distribution function ${f^{S}}$ needs to satisfy the following
kinetic moments:
\begin{equation}
M_{0}^{S}=\rho  \label{B1}
\end{equation}
\begin{equation}
\begin{array}{l}
M_{1x}^{S}=\rho {u_{x}} \\
M_{1y}^{S}=\rho {u_{y}}%
\end{array}
\label{A2}
\end{equation}
\begin{equation}
M_{2,0}^{S}=\frac{\rho }{2}(u_{x}^{2}+u_{y}^{2}+{n_{2}}RT)  \label{A3}
\end{equation}
\begin{equation}
\begin{array}{l}
M_{2xx}^{S}=\rho (u_{x}^{2}+RT) \\
M_{2xy}^{S}=\rho {u_{x}}{u_{y}} \\
M_{2yy}^{S}=\rho (u_{y}^{2}+RT)%
\end{array}
\label{A4}
\end{equation}
\begin{equation}
\begin{array}{l}
M_{3,1x}^{S}=\frac{{\rho {u_{x}}}}{2}(u_{x}^{2}+u_{y}^{2}+{n_{4}}RT)+(1-\Pr ){%
q_{x}} \\
M_{3,1y}^{S}=\frac{{\rho {u_{y}}}}{2}(u_{x}^{2}+u_{y}^{2}+{n_{4}}RT)+(1-\Pr ){%
q_{y}}%
\end{array}
\label{A5}
\end{equation}
\begin{equation}
\begin{array}{l}
M_{3xxx}^{S}=\rho {u_{x}}(u_{x}^{2}+3RT)+6n_{4}^{-1}(1-\Pr ){q_{x}} \\
M_{3xxy}^{S}=\rho {u_{y}}(u_{x}^{2}+RT)+2n_{4}^{-1}(1-\Pr ){q_{y}} \\
M_{3xyy}^{S}=\rho {u_{x}}(u_{y}^{2}+RT)+2n_{4}^{-1}(1-\Pr ){q_{x}} \\
M_{3yyy}^{S}=\rho {u_{y}}(u_{y}^{2}+3RT)+6n_{4}^{-1}(1-\Pr ){q_{y}}%
\end{array}
\label{A6}
\end{equation}
\begin{equation}
\begin{aligned} M_{4,2xx}^S &= \frac{\rho }{2}\left[ u_x^2(u_x^2 + u_y^2) +
n_4 R^2 T^2 + (n_7 u_x^2 + u_y^2)RT \right] \\ &\quad + 2n_4^{-1}(1 - \Pr)
(n_7 u_x q_x + u_y q_y) \\ M_{4,2xy}^S &= \frac{\rho }{2}\left[ u_x u_y
(u_x^2 + u_y^2) + n_6 u_x u_y RT \right] \\ &\quad + n_6 n_4^{-1}(1 - \Pr)
(u_x q_y + u_y q_x) \\ M_{4,2yy}^S &= \frac{\rho }{2}\left[ u_y^2 (u_x^2 +
u_y^2) + n_4 R^2 T^2 + (n_7 u_y^2 + u_x^2)RT \right] \\ &\quad + 2n_4^{-1}(1
- \Pr) (n_7 u_y q_y + u_x q_x) \end{aligned}  \label{A7}
\end{equation}
\begin{equation}
\begin{aligned} M_{4xxxx}^S &= \rho \left[ {u_x^2(u_x^2 + 6RT) +3{R^2}{T^2}}
\right] + 24n_4^{ - 1}(1 - \Pr ){u_x}{q_x}\\ M_{4xxxy}^S &= \rho
{u_x}{u_y}(u_x^2 + 3RT) + 6n_4^{ - 1}(1 - \Pr )({u_x}{q_y} + {u_y}{q_x})\\
M_{4xxyy}^S &= \rho \left[ {u_x^2u_y^2 + (u_x^2 + u_y^2)RT + {R^2}{T^2}}
\right] \\ &\quad + 4n_4^{ - 1}(1 - \Pr )({u_x}{q_x} + {u_y}{q_y})\\
M_{4xyyy}^S &= \rho {u_x}{u_y}(u_y^2 + 3RT) + 6n_4^{ - 1}(1 - \Pr
)({u_x}{q_y} + {u_y}{q_x})\\ M_{4yyyy}^S &= \rho \left[ {u_y^2(u_y^2 + 6RT)
+ 3{R^2}{T^2}} \right] + 24n_4^{ - 1}(1 - \Pr ){u_y}{q_y} \end{aligned}
\label{A8}
\end{equation}
\begin{equation}
\begin{aligned} M_{5,3xxx}^S &= \frac{\rho }{2}\left[ {u_x^3(u_x^2 + u_y^2) +
3{n_6}{u_x}{R^2}{T^2} + ({n_{11}}u_x^2 + 3u_y^2){u_x}RT} \right]\\ &\quad +
3n_4^{ - 1}(1 - \Pr )\left[ {{q_x}({n_{11}}u_x^2 + u_y^2 + 2{n_6}RT) +
2{u_x}{u_y}{q_y}} \right]\\ M_{5,3xxy}^S &= \frac{\rho }{2}\left[
{u_x^2{u_y}(u_x^2 + u_y^2) + {n_6}{u_y}{R^2}{T^2} + ({n_9}u_x^2 +
u_y^2){u_y}RT} \right]\\ &\quad + n_4^{ - 1}(1 - \Pr )\left[
{{q_y}({n_9}u_x^2 + 3u_y^2 + 2{n_6}RT) + 2{n_9}{u_x}{u_y}{q_x}} \right]\\
M_{5,3xyy}^S &= \frac{\rho }{2}\left[ {{u_x}u_y^2(u_x^2 + u_y^2) +
{n_6}{u_x}{R^2}{T^2} + ({n_9}u_y^2 + u_x^2){u_x}RT} \right]\\ &\quad + n_4^{
- 1}(1 - \Pr )\left[ {{q_x}({n_9}u_y^2 + 3u_x^2 + 2{n_6}RT) +
2{n_9}{u_x}{u_y}{q_y}} \right]\\ M_{5,3yyy}^S &= \frac{\rho }{2}\left[
{u_y^3(u_x^2 + u_y^2) + 3{n_6}{u_y}{R^2}{T^2} + ({n_{11}}u_y^2 +
3u_x^2){u_y}RT} \right]\\ &\quad + 3n_4^{ - 1}(1 - \Pr )\left[
{{q_y}({n_{11}}u_y^2 + u_x^2 + 2{n_6}RT) + 2{u_x}{u_y}{q_x}} \right],
\end{aligned}  \label{A9}
\end{equation}
\begin{equation}
\begin{aligned} M_{4,0}^S &= \frac{\rho }{2}\left[ {{{\left( {u_x^2 + u_y^2}
\right)}^2} + {n_2}{n_4}{R^2}{T^2} + 2{n_4}\left( {u_x^2 + u_y^2} \right)RT}
\right]\\ &\quad + 4(1 - \Pr )({u_x}{q_x} + {u_y}{q_y}). \end{aligned}
\label{A10}
\end{equation}
When $\Pr =1$, the above equations condense into the moment relations that $%
f^{eq}$ must satisfy to fully characterize first- and second-order viscous
stress and heat flux.


\bibliographystyle{jfm}
\bibliography{jfm-instructions-fs}

\end{document}